\documentclass[conference]{IEEEtran}
\IEEEoverridecommandlockouts
\usepackage{cite}
\usepackage{amsmath,amssymb,amsfonts}
\usepackage{graphicx}
\usepackage{textcomp}
\usepackage{xcolor}
\usepackage{subcaption}
\usepackage{algorithm}
\usepackage{algpseudocode}
\usepackage{multirow}
\usepackage{hyperref}
\usepackage{xspace}
\usepackage{placeins}
\usepackage{booktabs}
\usepackage{multirow}
\usepackage{rotating}
\usepackage{enumitem}

\def\BibTeX{{\rm B\kern-.05em{\sc i\kern-.025em b}\kern-.08em
    T\kern-.1667em\lower.7ex\hbox{E}\kern-.125emX}}

\begin{document}

\title{Parameterized Task Graph Scheduling Algorithm for Comparing Algorithmic Components\thanks{This work was supported in part by Army Research Laboratory under Cooperative Agreement W911NF-17-2-0196. The authors acknowledge the Center for Advanced Research Computing (CARC) at the University of Southern California for providing computing resources that have contributed to the research results reported within this publication. URL: \url{https://carc.usc.edu}.}}

\author{\IEEEauthorblockN{1\textsuperscript{st} Jared Coleman}
\IEEEauthorblockA{\textit{Thomas Lord Department of Computer Science} \\
\textit{University of Southern California}\\
Los Angeles, California, USA \\
jaredcol@usc.edu}
\and
\IEEEauthorblockN{2\textsuperscript{nd} Ravi Vivek Agrawal}
\IEEEauthorblockA{\textit{Thomas Lord Department of Computer Science} \\
\textit{University of Southern California}\\
Los Angeles, California, USA \\
ravivive@usc.edu}
\and
\IEEEauthorblockN{3\textsuperscript{rd} Ebrahim Hirani}
\IEEEauthorblockA{\textit{Thomas Lord Department of Computer Science} \\
\textit{University of Southern California}\\
Los Angeles, California, USA \\
ehirani@usc.edu}
\and
\IEEEauthorblockN{4\textsuperscript{th} Bhaskar Krishnamachari}
\IEEEauthorblockA{\textit{Ming Hsieh Department of Electrical and Computer Engineering} \\
\textit{University of Southern California}\\
Los Angeles, California, USA \\
bkrishna@usc.edu}
}

\maketitle

\begin{abstract}
Scheduling distributed applications modeled as directed, acyclic task graphs to run on heterogeneous compute networks is a fundamental (NP-Hard) problem in distributed computing for which many heuristic algorithms have been proposed over the past decades.
Many of these algorithms fall under the \textit{list-scheduling} paradigm, whereby the algorithm first computes priorities for the tasks and then schedules them greedily to the compute node that minimizes some cost function.
Thus, many algorithms differ from each other only in a few key components (e.g., the way they prioritize tasks, their cost functions, where the algorithms consider inserting tasks into a partially complete schedule, etc.).
In this paper, we propose a generalized list-scheduling algorithm that allows mixing and matching different task prioritization and greedy node selection schemes to produce 72 unique algorithms.
We benchmark these algorithms on four datasets to study the individual effect of different algorithmic components on performance and runtime.
\end{abstract}


\section{Introduction}
Task graph scheduling is a fundamental problem in distributed computing.
Essentially, the goal is to assign computational tasks to different compute nodes in such a way that minimizes/maximizes some performance metric (e.g., total execution time, energy consumption, throughput, etc.).
In this paper, we will focus on the task scheduling problem concerning \textit{heterogeneous} task graphs and compute networks with the objective of \textit{minimizing makespan} (total execution time) under the \textit{related machines} model\footnote{In the related machines model, if the same task executes faster one some compute node $n_1$ than on node $n_2$, then $n_1$ must execute \textit{all} tasks faster than $n_2$ ($n_1$ is strictly faster than $n_2$). 
The related machines model as it pertains to the task scheduling problem we study in this paper is described further in Section~\ref{sec:problem}.}.
As this problem is NP-hard~\cite{theory:npcomplete} and also not polynomial-time approximable within a constant factor~\cite{theory:inapproximable}, many heuristic algorithms have been proposed.
One of the most popular paradigms for heuristic scheduling algorithms is the \textit{list-scheduling} paradigm.
Essentially, list-scheduling algorithms involve the following steps:
\begin{enumerate}
    \item Compute a priority for each task such that every task has a higher priority than its dependents.
    \item Greedily schedule tasks in order of their computed priorities (from highest to lowest) to run on the node that minimizes/maximizes some predefined cost function.
\end{enumerate}
Thus, many list scheduling algorithms differ only in a few algorithmic components (e.g., their prioritization functions, their cost functions, where the algorithms consider inserting tasks into an existing schedule, etc.).
In this paper, we propose a general parametric scheduler (extending SAGA~\cite{framework_repo}, an open-source python library for comparing task scheduling algorithms) that allows us to mix-and-match different algorithmic components and evaluate how they individually contribute to a list-scheduling algorithm's performance and runtime.
Interestingly, we find that many \textit{new} algorithms (composed of previously unstudied combinations of algorithmic components) are pareto-optimal with respect to performance and runtime.
We also report the average effects that both individual components \textit{and} combinations of different components have on performance and runtime, presenting evidence that the way algorithmic components interact with each other is problem-dependent (i.e., depends on the task graph structure, whether or not the application is communication or computation heavy, etc.).

\subsection{Problem Definition}\label{sec:problem}
Let us denote the task graph as $G = (T,D)$, where $T$ is the set of tasks and $D$ contains the directed edges or dependencies between these tasks. 
An edge $(t,t^\prime) \in D$ implies that the output from task $t$ is required input for task $t^\prime$. 
Thus, task $t^\prime$ cannot start executing until it has received the output of task $t$. 
This is often referred to as a precedence constraint. 
For a given task $t \in T$, its compute cost is represented by $c(t) \in \mathbb{R}^+$ and the size of the data exchanged between two dependent tasks, $(t, t^\prime) \in D$, is $c(t, t^\prime) \in \mathbb{R}^+$. 
Let $N = (V, E)$ denote the compute node network, where $N$ is a complete undirected graph. 
$V$ is the set of nodes and $E$ is the set of edges. 
The compute speed of a node $v \in V$ is $s(v) \in \mathbb{R}^+$ and the communication strength between nodes $(v, v^\prime) \in E$ is $s(v,v^\prime) \in \mathbb{R}^+$. 
Under the \textit{related machines} model~\cite{graham}, the execution time of a task $t \in T$ on a node $v \in V$ is $ \frac{c(t)}{s(v)}$, and the data communication time between tasks $(t, t^\prime) \in D$ from node $v$ to node $v^\prime$ (i.e., $t$ executes on $v$ and $t^\prime$ executes on $v^\prime$) is $\frac{c(t, t^\prime)}{s(v, v^\prime)}$.

The goal is to schedule the tasks on different compute nodes in such a way that minimizes the makespan (total execution time) of the task graph.
Let $\mathcal{A}$ denote a task scheduling algorithm.
For a given a problem instance $(N, G)$ which represents a network/task graph pair, a \textit{schedule} is a set of tuples of the form $(t, v, r, e)$ where $t \in T$, $v \in V$ is the node on which task $t$ is scheduled to run, $r \in \mathbb{R}^+$ is the start time, and $e \in \mathbb{R}^+$ is the end time.
A valid schedule $S$ must satisfy the following properties
\begin{itemize}
    \item All tasks must be scheduled exactly once:
    \begin{gather*}
        \forall t \in T, \exists (t, v, r, e) \in S \\
        \forall (t, v, r, e), (t^\prime, v^\prime, r^\prime, e^\prime) \in S,t = t^\prime \\
        \hspace{8em} \Rightarrow (v, r, e) = (v^\prime, r^\prime, e^\prime)
    \end{gather*}
    \item All tasks must have valid start and end times:
    $$
        \forall (t, v, r, e) \in S, e - r = \frac{c(t)}{s(v)}
    $$
    \item Only one task can be scheduled on a node at a time (i.e., their start/end times cannot overlap):
    $$
        \forall (t, v, r, e), (t^\prime, v^\prime, r^\prime, e^\prime) \in S, e \leq r^\prime \lor e^\prime \leq r
    $$
    \item A task cannot start executing until all of its dependencies have finished executing and their outputs have been received at the node on which the task is scheduled:
    \begin{gather*}
        \forall (t, v, r, e), (t^\prime, v^\prime, r^\prime, e^\prime) \in S, (t, t^\prime) \in D \\
        \hspace{8em} \Rightarrow e + \frac{c(t, t^\prime)}{s(v, v^\prime)} \leq r^\prime
    \end{gather*}
\end{itemize}

\begin{figure*}[h!]
    \begin{subfigure}{0.5\textwidth}
        \centering
        \includegraphics[width=\linewidth]{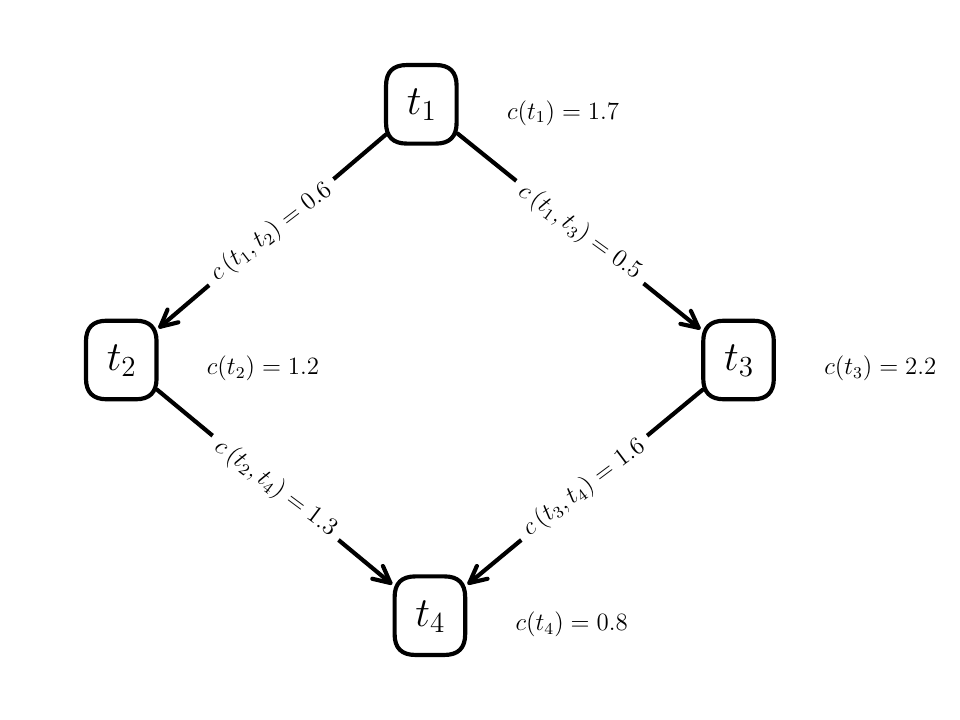}
        \caption{Task Graph}
        \label{fig:example:task_graph}
    \end{subfigure}%
    \begin{subfigure}{0.5\textwidth}
        \centering
        \includegraphics[width=\linewidth]{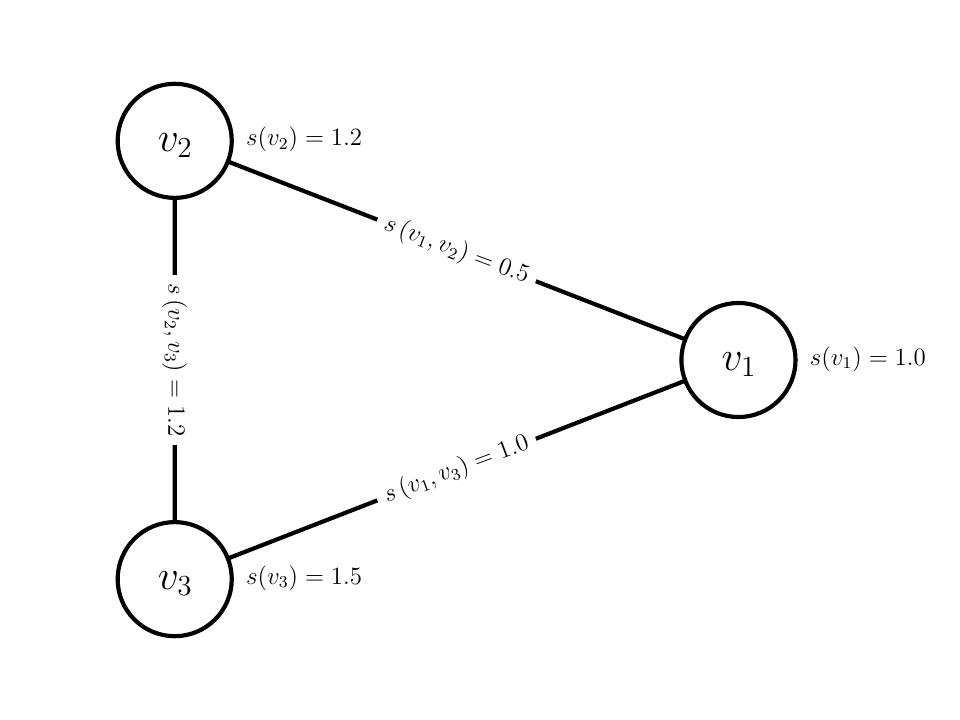}
        \caption{Network}
        \label{fig:example:network}
    \end{subfigure}%
    
    \vspace{0.25cm}%
    
    \begin{subfigure}{\textwidth}
        \centering
        \includegraphics[width=0.8\linewidth]{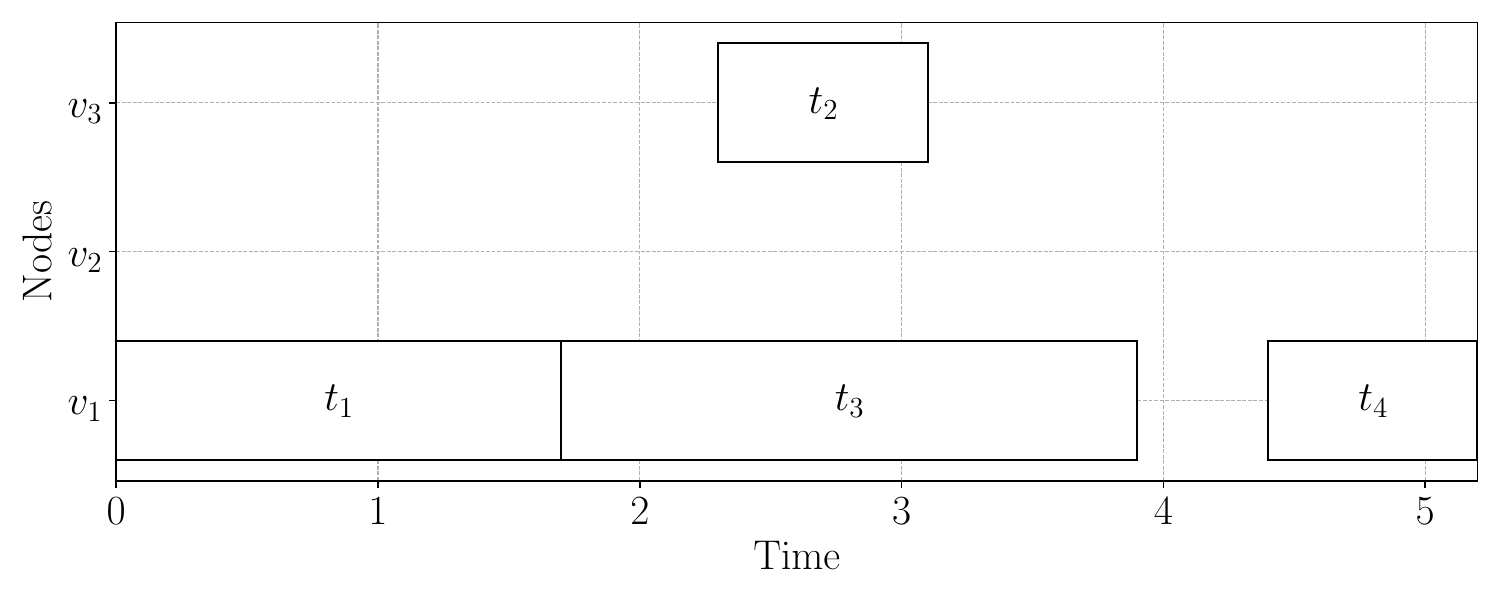}
        \caption{Schedule}
        \label{fig:example:heft}
    \end{subfigure}
    \caption{Example problem instance and schedule.}
    \label{fig:example}
\end{figure*}

Figure~\ref{fig:example} depicts an example problem instance (task graph and network) and solution (schedule).
We define the makespan of a schedule $S$ as the time at which the last task finishes executing:
$$
    m(S) = \max_{(t, v, r, e) \in S} e
$$

Because the problem of minimizing makespan is NP-hard for this model~\cite{theory:npcomplete}, many heuristic algorithms have been proposed.
Traditionally, these heuristic algorithms are evaluated on a set of problem instances and compared against other algorithms based on their makespan ratio, which for a given problem instance $(N, G)$ is the makespan of the schedule produced by the algorithm divided by the minimum makespan of the schedules produced by the baseline algorithms.
Let $S_{\mathcal{A},N,G}$ denote the schedule produced by algorithm $\mathcal{A}$ on problem instance $(N, G)$.
Then the makespan ratio of an algorithm $\mathcal{A}$ against a set of baseline algorithms $\mathcal{A}_1, \mathcal{A}_2, \ldots$ for a problem instance $(N, G)$ can be written
\begin{align*}
    \frac{m\left(S_{\mathcal{A},N,G}\right)}{\min\left\{
        m\left(S_{\mathcal{A}_1,N,G}\right), m\left(S_{\mathcal{A}_2,N,G}\right), m\left(S_{\mathcal{A}_3,N,G}\right), \ldots
    \right\}} .
\end{align*}

Makespan ratio is a commonly used concept in the task scheduling literature~\cite{scheduler:heft,compare:list_vs_cluster}.
It is common to measure the makespan ratios of an algorithm (against a set of baseline algorithms) for a dataset of problem instances.
System designers use this technique, called \textit{benchmarking}, to decide which algorithm(s) best support(s) their application.
In this paper, we study how algorithmic components affect both the makespan \textit{and} runtime (amount of time an algorithm takes to produce a schedule).
Let $r_{\mathcal{A}}(N, G)$ denote the amount of time it takes to produce a schedule (on a particular system) using algorithm $\mathcal{A}$.
Then, the runtime ratio of an algorithm $\mathcal{A}$ against a set of baseline algorithms $\mathcal{A}_1, \mathcal{A}_2, \ldots$ for a problem instance $(N, G)$ can be written
\begin{align*}
\frac{r_{\mathcal{A}}\left(N,G\right)}{\min\left\{
        r_{\mathcal{A}_1}\left(N,G\right), r_{\mathcal{A}_2}\left(N,G\right), r_{\mathcal{A}_3}\left(N,G\right), \ldots
    \right\}} .
\end{align*}
In reality, the actual runtime of an algorithm even on a particular instance is not deterministic (due, for example, to background processes running on the system).
Thus, the runtime ratios reported in this paper should be interpreted as \textit{estimated} runtime ratios, rather than absolute deterministic values (unlike the makespan ratios).

\subsection{Contributions}
The main contributions of this work are:
\begin{itemize}
    \item Proposes an open-source generalized parametric list scheduling algorithm that allows mixing and matching of different algorithmic components.
    \item Reports benchmarking results for 72 algorithms produced by combining five different algorithmic components on 20 publicly available datasets.
    \item Reports both individual and combined effects of different algorithmic components on average performance and runtime across all datasets.
    \item Reports both individual and combined effects of different algorithmic components on performance and runtime for each individual dataset.
\end{itemize}
The rest of this paper is organized as follows.
In Section~\ref{sec:related} we survey related work in comparing task scheduling algorithms and algorithmic components of such algorithms.
Then, in Section~\ref{sec:solution} we present our methodology for comparing different algorithmic components and the generalized parametric list scheduling algorithm for doing so.
In Section~\ref{sec:results} we report benchmarking results for 72 algorithms created by combining five different types of algorithmic components.
We study the effects that individual components have on performance and runtime and their interactions with other components.
Finally, in Section~\ref{sec:conclusion} we conclude the paper with a summary of the contributions and directions for future work.

\section{Related Work}\label{sec:related}
Existing approaches to comparing task scheduling algorithms mostly involve benchmarking, whereby a set of algorithms is evaluated on one or more datasets and various performance metrics are reported for comparison. One of the most common metrics (and the one used in this paper) is the makespan ratio (also called schedule length ratio) which is the makespan, or total execution time, of the schedule produced by an algorithm for a given problem instance normalized by the \textit{minimum} makespan produced by any of the algorithms being evaluated. Other metrics often reported include speedup (how many times larger the makespan would be if the algorithm scheduled all tasks to run on a single node), efficiency (average speedup per node), frequency that the algorithm is the best algorithm among those being evaluated, and slack (a measurement of schedule robustness)~\cite{compare:benchmarking_hetero}.

Among the many notable benchmarking efforts in the literature, one study~\cite{compare:benchmarking_schedulers} benchmarks 15 scheduling algorithms and even discusses many of the algorithmic components that we consider in this paper, including Assigning Priorities to Nodes (tasks), Insertion vs. Non-Insertion, and Critical-Path-Based vs. non Critical-Path-Based.
They evaluate on different task graphs: Peer Set Graphs and Random Graphs without studying the effects of these components on performance.
Another evaluation~\cite{scheduler:eleven} evaluates eleven algorithms for independent non-communicating tasks with uniformly random costs and heterogeneous compute nodes with uniformly random speeds and yet another~\cite{compare:benchmarking_hetero} evaluates six algorithms on randomly generated application graphs and real-world application graphs (FFT, Gaussian Elimination, Montage, and Epigenomics Scientific Workflow).
Metaheuristic algorithms like simulated annealing and genetic algorithms have also been evaluated using a similar methodology~\cite{compare:metaheuristics}.
A comprehensive evaluation of 31 list scheduling algorithms produced by combining different algorithmic components has been conducted~\cite{compare:list_vs_cluster} (the same paper also presents results for \textit{cluster scheduling} which group tasks into clusters before scheduling them), reporting useful benchmarking results but do nothing regarding effect that individual algorithmic components have on performance.
Unfortunately, many of the existing comparison efforts are also difficult to reproduce since the datasets typically not publicly available and/or the scheduler implementations are not open-sourced.

To the best of our knowledge, this work is the first to evaluate the performance of all possible combinations of a set of algorithmic components and use the results to study the individual and combined effects these components have on the performance (makespan ratio) and runtime of list-scheduling algorithms.

\section{A Generalized List-Scheduling Algorithm}\label{sec:solution}
To study the effects of individual algorithmic components in list-scheduling algorithms, we extended \textsc{SAGA}~\cite{framework_repo} (a python library for comparing task graph scheduling algorithms) with a generalized parametric scheduling algorithm that allows users to specify the following algorithmic components:
\begin{itemize}
    \item Priority Function: Function used to determine the order in which to schedule tasks.
    \item Comparison Function: Function used to determine which node to schedule a task to.
    \item Insertion-based vs. append-only scheduling: Insertion-based algorithms insert tasks into the earliest sufficiently large window (on a particular node) in an existing schedule. Append-only algorithms only consider scheduling tasks \textit{after} the last currently scheduled task finishes on a particular node.
    \item Critical path reservation vs. no reservation: Whether or not the algorithm commits to scheduling all tasks on the critical path --- the longest (with respect to node/edge weights) chain of tasks in the task graph --- on the fastest compute node~\footnote{This is consistent with the original definition~\cite{scheduler:heft} of critical path and critical path reservation for the task scheduling problem we study in this paper (fully heterogeneous, related machines model) but may not be for other task scheduling varaints (e.g., for the unrelated machines model)}.
    \item Sufferage vs. no-sufferage consideration: Sufferage schedulers~\cite{scheduler:sufferage} consider the top 2 highest-priority task in each iteration, choosing to schedule the one for which scheduling on the second-best node would cause the greatest detriment (with respect to the comparison function).
\end{itemize}
The algorithm works first by using the configured priority function to determine the order in which to schedule tasks.
The priority function takes the problem instance (network/task graph pair) as input and returns a sequence of tasks.
In each iteration, the algorithm uses the comparison function and whether or not the algorithm is using insertion-based or append-only scheduling to decide which node it should schedule a task on.
In insertion-based scheduling, the scheduler finds, for each node, the earliest window of time for which the node is idle and the window is large enough for the task to execute (taking into account communication delays for dependency data transfer).
In the append-only scheme, the scheduling algorithm only considers scheduling tasks after the last currently scheduled task on a node finishes executing.
Once the algorithm computes the candidate start/end times for the task on each candidate node, it uses the comparison method to choose which node to select.
The comparison function may consider the candidate task's start time, end time, and/or other information to make this decision.
If the algorithm is a sufferage scheduler, then the comparison function is used identify the best \textit{and} the second-best node on which to schedule the task and then used again to quantify the candidate task's \textit{sufferage}: the difference in quality between scheduling on the second-best and best node.
The algorithm then chooses to schedule the task the highest \textit{sufferage} value (the idea being that not scheduling this task on its preferred node is more detrimental) and returns the other task to the queue.

The priority and comparison functions are very general algorithmic components for which there are many possible implementations.
In this paper, we consider the following implementations of each algorithmic component:
\begin{itemize}
    \item Priority Function (initial\_priority)
    \begin{itemize}
        \item UpwardRank: the priority function used in the HEFT scheduling algorithm~\cite{scheduler:heft}
        \item CPoPRank: the priority function used in the CPoP scheduling algorithm~\cite{scheduler:heft}
        \item ArbitraryTopological: an arbitrary topological sort of the task graph
    \end{itemize}
    \item Comparison Function (compare)
    \begin{itemize}
        \item EFT: (Earliest Finish Time): schedules tasks to the node on which the task can finish the soonest
        \item EST (Earliest Start Time): schedules tasks to the node on which the task can start the soonest
        \item Quickest: schedules tasks to the node on which it executes in the least amount of time
    \end{itemize}
    \item Insertion-based vs. append-only scheduling (append\_only = True or False)
    \item Critical path reservation vs. no reservation (critical\_path = True or False)
    \item Sufferage vs. no-sufferage consideration (sufferage = True or False)
\end{itemize}
Algorithms~\ref{alg:eft},\ref{alg:est}, and~\ref{alg:quickest} implement the EFT, EST, and Quickest comparison functions, respectively.
Algorithms~\ref{alg:append-only} and~\ref{alg:insertion-based} implement functionality for append-only and insertion-based scheduling.
Implementations for the priority functions are not included.
Details for the UpwardRank and CPoPRank priority functions can be found in the original paper they were proposed in~\cite{scheduler:heft}.
Finally, Algorithm~\ref{alg:scheduler} represents the full parametric scheduling algorithm and uses the functions defined in Algorithms 1-5.

\begin{algorithm*}
\caption{Earliest Finish Time Compare Algorithm}\label{alg:eft}
\begin{algorithmic}[1]
\Function{EFT}{$(start, end)$, $(start^\prime, end^\prime)$}
    \Statex \textbf{inputs}: $(start, end) \in \mathbb{R}^2$ \Comment{start/end time of first candidate task}
    \Statex \phantom{\textbf{inputs}: }$(start^\prime, end^\prime) \in \mathbb{R}^2$ \Comment{start/end time of second candidate task}
    \State \Return $end - end^\prime$ \Comment{$< 0$ iff the first task is better (finishes earlier) than the second task}
\EndFunction
\end{algorithmic}
\end{algorithm*}%
\begin{algorithm*}
\caption{Earliest Start Time Compare Algorithm}\label{alg:est}
\begin{algorithmic}[1]
\Function{EST}{$(start, end)$, $(start^\prime, end^\prime)$}
    \Statex \textbf{inputs}: $(start, end) \in \mathbb{R}^2$ \Comment{start/end time of first candidate task}
    \Statex \phantom{\textbf{inputs}: }$(start^\prime, end^\prime) \in \mathbb{R}^2$ \Comment{start/end time of second candidate task}
    \State \Return $start - start^\prime$ \Comment{$< 0$ iff the first task is better (starts earlier) than the second task}
\EndFunction
\end{algorithmic}
\end{algorithm*} %
\begin{algorithm*}
\caption{Quickest Execution Time Compare Algorithm}\label{alg:quickest}
\begin{algorithmic}[1]
\Function{EST}{$(start, end)$, $(start^\prime, end^\prime)$}
    \Statex \textbf{inputs}: $(start, end) \in \mathbb{R}^2$ \Comment{start/end time of first candidate task}
    \Statex \phantom{\textbf{inputs}: }$(start^\prime, end^\prime) \in \mathbb{R}^2$ \Comment{start/end time of second candidate task}
    \State \Return $(end - start) - (end^\prime - start^\prime)$ \Comment{$< 0$ iff the first task is better (executes faster) than the second task}
\EndFunction
\end{algorithmic}
\end{algorithm*}

\begin{algorithm*}
\caption{Append-only available window finding algorithm}\label{alg:append-only}
\begin{algorithmic}[1]
\Function{GetOpenWindowAppendOnly}{$u$, $t$}
    \State Let $est$ be the finish time of the last task scheduled on node $u$ ($0$ if no tasks scheduled on $u$)
    \State Let $dat$ be the minimum time at which all data from task $t$ dependencies can be available on node $u$
    \State $start \gets \max\{est, dat\}$
    \State $end \gets start + \frac{c(t)}{s(u)}$
    \State \Return $(start, end)$
\EndFunction
\end{algorithmic}
\end{algorithm*}

\begin{algorithm*}
\caption{Insertion-based available window finding algorithm}\label{alg:insertion-based}
\begin{algorithmic}[1]
\Statex
\Function{GetOpenWindowInsertionBased}{$u$, $t$}
    \State $T \gets $ list of tasks scheduled on node $u$ in order of start time
    \If {$T$ is empty}
        \Return $\Call{GetOpenWindowAppendOnly}(u,t)$
    \EndIf
    \State Let $dat$ be the minimum time at which all data from task $t$ dependencies can be available on node $u$
    \For{task $t^\prime$ scheduled on node $u$} \Comment{iterate in order of task start time}
        \State Let $est$ be the time that task $t^\prime$ finishes
        \State $start \gets \max\{est, dat\}$
        \State $end \gets start + \frac{c(t)}{s(u)}$
        \State \textbf{if} $end \leq$ start time of next task in $T$ \textbf{or} $t$ is the last task in $T$ \textbf{then} \Return $(start, end)$
    \EndFor
\EndFunction
\end{algorithmic}
\end{algorithm*}

\begin{algorithm*}
\caption{Generalized Parametric Scheduling Algorithm}\label{alg:scheduler}
\begin{algorithmic}[1]
    \Statex \textbf{parameters}: $\textsc{GetPriority} \in \{ \textsc{UpwardRanking}, \textsc{CPoPRanking}, \textsc{ArbitraryTopological} \}$
    \Statex \phantom{\textbf{parameters}: }$\textsc{Compare} \in \{ \textsc{EFT}, \textsc{EST}, \textsc{Quickest} \}$
    \Statex \phantom{\textbf{parameters}: }$append\_only \in \{ 0, 1 \}$
    \Statex \phantom{\textbf{parameters}: }$sufferage \in \{ 0, 1 \}$
    \Statex \phantom{\textbf{parameters}: }$critical\_path \in \{ 0, 1 \}$
    \State Initialize empty schedule $S$
    \State Compute priority for each task in $G$ using Priority Function
    \State Sort tasks in descending order of priority
    \If {$append\_only$}
        \State Let $\textsc{GetWindow} \gets \textsc{GetOpenWindowAppendOnly}$
    \Else
        \State Let $\textsc{GetWindow} \gets \textsc{GetOpenWindowInsertionBased}$
    \EndIf
    \While{unscheduled tasks exist}
        \State initialize $best$ and $second\_best$ to arbitrary nodes
        \State Let $t$ be the unscheduled task with highest priority
        \For {$u \in V$}
            \If {$\Call{Compare}{\textsc{GetWindow}(t, u), \textsc{GetWindow}(t, best)} < 0$}
                \State $second\_best \gets best$
                \State $best \gets u$
            \ElsIf{$\Call{Compare}{\textsc{GetWindow}(t, u), \textsc{GetWindow}(t, second\_best)} < 0$}
                \State $second\_best \gets u$
            \EndIf
        \EndFor
        \If {$sufferage$ \textbf{and} there are at least two unscheduled tasks}
            \State initialize $best^\prime$ and $second\_best^\prime$ to arbitrary nodes
            \State Let $t^\prime$ be the unscheduled task with second-highest priority
            \For {$u \in V$}
                \If {$\Call{Compare}{\textsc{GetWindow}(t^\prime, u), \textsc{GetWindow}(t^\prime, best^\prime)} < 0$}
                    \State $second\_best^\prime \gets best^\prime$
                    \State $best^\prime \gets u$
                \ElsIf{$\Call{Compare}{\textsc{GetWindow}(t^\prime, u), \textsc{GetWindow}(t^\prime, second\_best^\prime)} < 0$}
                    \State $second\_best^\prime \gets u$
                \EndIf
            \EndFor
            \State $sufferage\_t \gets $\Call{Compare}{$(t, second\_best), (t, best)$}
            \State $sufferage\_t^\prime \gets $\Call{Compare}{$(t^\prime, second\_best^\prime), (t^\prime, best^\prime)$}
            \If { $sufferage\_t^\prime > sufferage\_t$}
                \State $t, best \gets t^\prime, best^\prime$
            \EndIf
        \EndIf
        \State Add $(t, start, end, best)$ to $S$
    \EndWhile
    \State \Return $S$
\end{algorithmic}
\end{algorithm*}

\begin{figure*}[!h]
    \centering

    \begin{subfigure}[b]{0.48\textwidth}
        \centering
        \includegraphics[width=\textwidth]{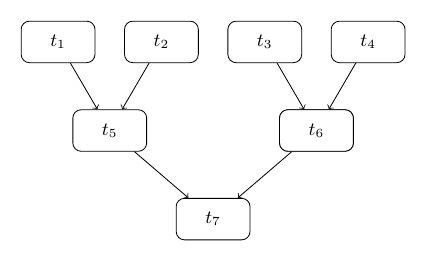}
        \caption{Example task graph structure for the in-trees datasets.}
        \label{fig:task_graphs:in_trees}
    \end{subfigure}%
    \hfill
    \begin{subfigure}[b]{0.48\textwidth}
        \centering
        \includegraphics[width=\textwidth]{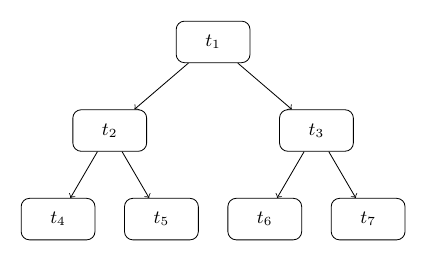}
        \caption{Example task graph structure for the out-trees datasets.}
        \label{fig:task_graphs:out_trees}
    \end{subfigure}%
    \hfill
    \begin{subfigure}[b]{0.48\textwidth}
        \centering
        \includegraphics[width=\textwidth]{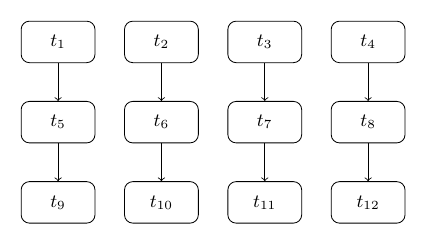}
        \caption{Example task graph structure for the chains datasets.}
        \label{fig:task_graphs:chains}
    \end{subfigure}%
    \hfill
    \begin{subfigure}[b]{0.48\textwidth}
        \centering
        \includegraphics[width=\textwidth]{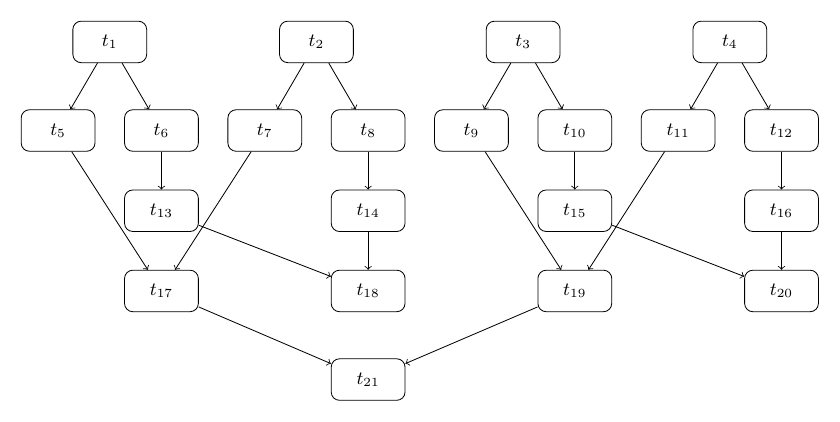}
        \caption{Example task graph structure for the cycles datasets.}
        \label{fig:task_graphs:cycles}
    \end{subfigure}%
    \caption{Example task graphs for each of the datasets evaluated.}
    \label{fig:task_graphs}
\end{figure*}

The combinations of these algorithmic components allow for $72$ unique scheduling algorithms.
We evaluated each of these on twenty datasets based on four task graph structures (chains, in\_trees, and out\_trees, cycles) and five communication-to-computation ratios, or CCRs, (CCR $1/5$, $1/2$, $1$, $2$, $5$) for each.
Each in\_trees, out\_trees, and chains datasets consists of $100$ randomly generated network/task graph pairs following a common methodology used in the literature~\cite{data:random_graphs}.
In-trees and out-trees are generated with between $2$ and $4$ levels (chosen uniformly at random), a branching factor of either $2$ or $3$ (chosen uniformly at random), and node/edge-weights drawn from a clipped Gaussian distribution (mean: $1$, standard deviation: $1/3$, min: $0$, max: $2$).
Parallel chains task graphs are generated with between $2$ and $5$ parallel chains (chosen uniformly at random) of length between $2$ and $5$ (chosen uniformly at random) and node/edge-weights drawn from the same clipped Gaussian distribution.
Randomly weighted networks are complete graphs with between $3$ and $5$ nodes (chosen uniformly at random) and node/edge-weights drawn from the same clipped Gaussian distribution.
Finally, the network communication strengths are scaled to achieve the desired CCR of either $1/5$, $1/2$, $1$, $2$, or $5$.
The cycles dataset is based on the cycles scientific workflow~\cite{data:cycles} and represents a multi-crop, multi-year agro-ecosystem model for simulating crop production.
For each workflow, the runtime of each task, input/output sizes in bytes, and speedup factor (compute speed) for each machine are available from public execution trace information\footnote{\href{https://github.com/wfcommons/pegasus-instances}{https://github.com/wfcommons/pegasus-instances}, \href{https://github.com/wfcommons/makeflow-instances}{https://github.com/wfcommons/makeflow-instances}}.
We set communication strengths to be homogeneous so that the average CCR is $\frac{1}{5}, \frac{1}{2}, 1, 2, $ or $5$ (resulting in five different datasets).
Figure~\ref{fig:task_graphs} shows an example in\_trees, out\_trees, chains, and cycles task graph.

\section{Results}\label{sec:results}
In this section we discuss the results of running each of the $72$ schedulers on $20$ datasets ($4$ dataset types and $5$ CCRs).
Each dataset consists of $100$ problem instances.
All schedulers that are \textit{pareto optimal} for at least one of the evaluated datasets are listed in Table~\ref{tab:pareto}.
There are $24$ such schedulers ($48$ of the $72$ total schedulers evaluated were strictly dominated by at least one other scheduler for every dataset).

\begin{table*}[!h]
\centering
\begin{tabular}{|l|lrlrr|}
\hline
\hfill component & initial\_priority & append\_only & compare & critical\_path & sufferage \\
scheduler & & & & &  \\
\hline
EFT\_Ins\_AT & ArbitraryTopological & False & EFT & False & False \\
EFT\_Ins\_AT\_Suf & ArbitraryTopological & False & EFT & False & True \\
EST\_Ins\_AT & ArbitraryTopological & False & EST & False & False \\
EST\_Ins\_AT\_Suf & ArbitraryTopological & False & EST & False & True \\
EST\_Ins\_CP\_AT & ArbitraryTopological & False & EST & True & False \\
MCT~\cite{scheduler:eleven} & ArbitraryTopological & True & EFT & False & False \\
Sufferage~\cite{scheduler:sufferage} & ArbitraryTopological & True & EFT & False & True \\
EFT\_App\_CP\_AT & ArbitraryTopological & True & EFT & True & False \\
EST\_App\_AT & ArbitraryTopological & True & EST & False & False \\
EST\_App\_AT\_Suf & ArbitraryTopological & True & EST & False & True \\
MET~\cite{scheduler:eleven} & ArbitraryTopological & True & Quickest & False & False \\
EFT\_Ins\_CR & CPoPRanking & False & EFT & False & False \\
EFT\_Ins\_CR\_Suf & CPoPRanking & False & EFT & False & True \\
EST\_Ins\_CR & CPoPRanking & False & EST & False & False \\
EST\_Ins\_CR\_Suf & CPoPRanking & False & EST & False & True \\
EFT\_App\_CR\_Suf & CPoPRanking & True & EFT & False & True \\
HEFT~\cite{scheduler:heft} & UpwardRanking & False & EFT & False & False \\
EFT\_Ins\_UR\_Suf & UpwardRanking & False & EFT & False & True \\
EST\_Ins\_UR & UpwardRanking & False & EST & False & False \\
EST\_Ins\_UR\_Suf & UpwardRanking & False & EST & False & True \\
EFT\_App\_UR & UpwardRanking & True & EFT & False & False \\
EFT\_App\_UR\_Suf & UpwardRanking & True & EFT & False & True \\
EST\_App\_UR & UpwardRanking & True & EST & False & False \\
EST\_App\_UR\_Suf & UpwardRanking & True & EST & False & True \\
\hline
\end{tabular}
\caption{All scheduling algorithms that were pareto-optimal for at least one evaluated dataset ($24$ of $72$ schedulers). All other schedulers ($48$ of $72$) were strictly dominated by another algorithm in every dataset.\vspace{2em}}
\label{tab:pareto}
\end{table*}

\begin{figure*}
    \begin{subfigure}{\textwidth}
        \centering
        \includegraphics[width=\linewidth]{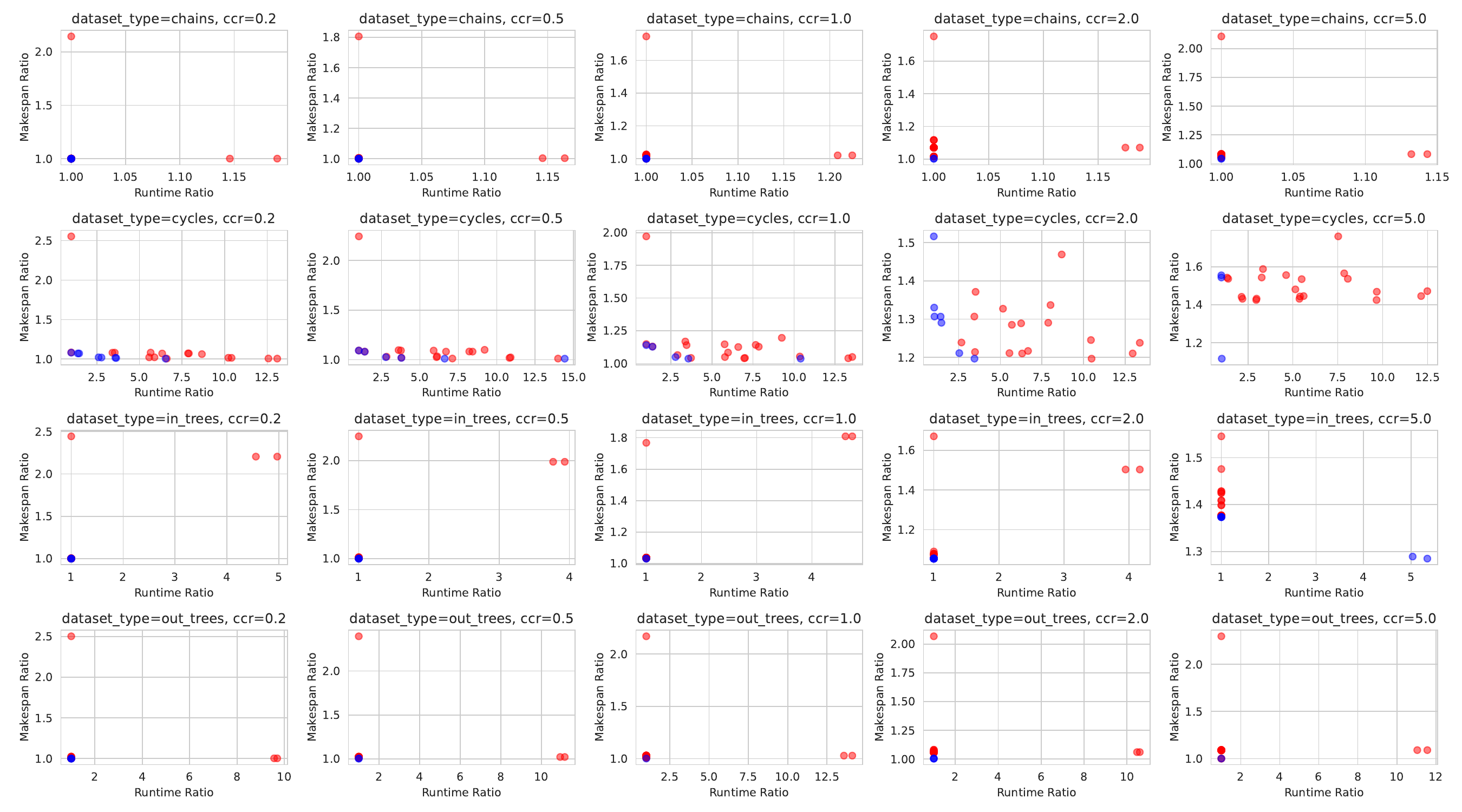}
        \caption{Markers represent scheduling algorithms that are pareto-optimal for at least one of the evaluated datasets.}\label{fig:pareto:scatter}
    \end{subfigure}
    \begin{subfigure}{\textwidth}
        \centering
        \includegraphics[width=0.8\linewidth]{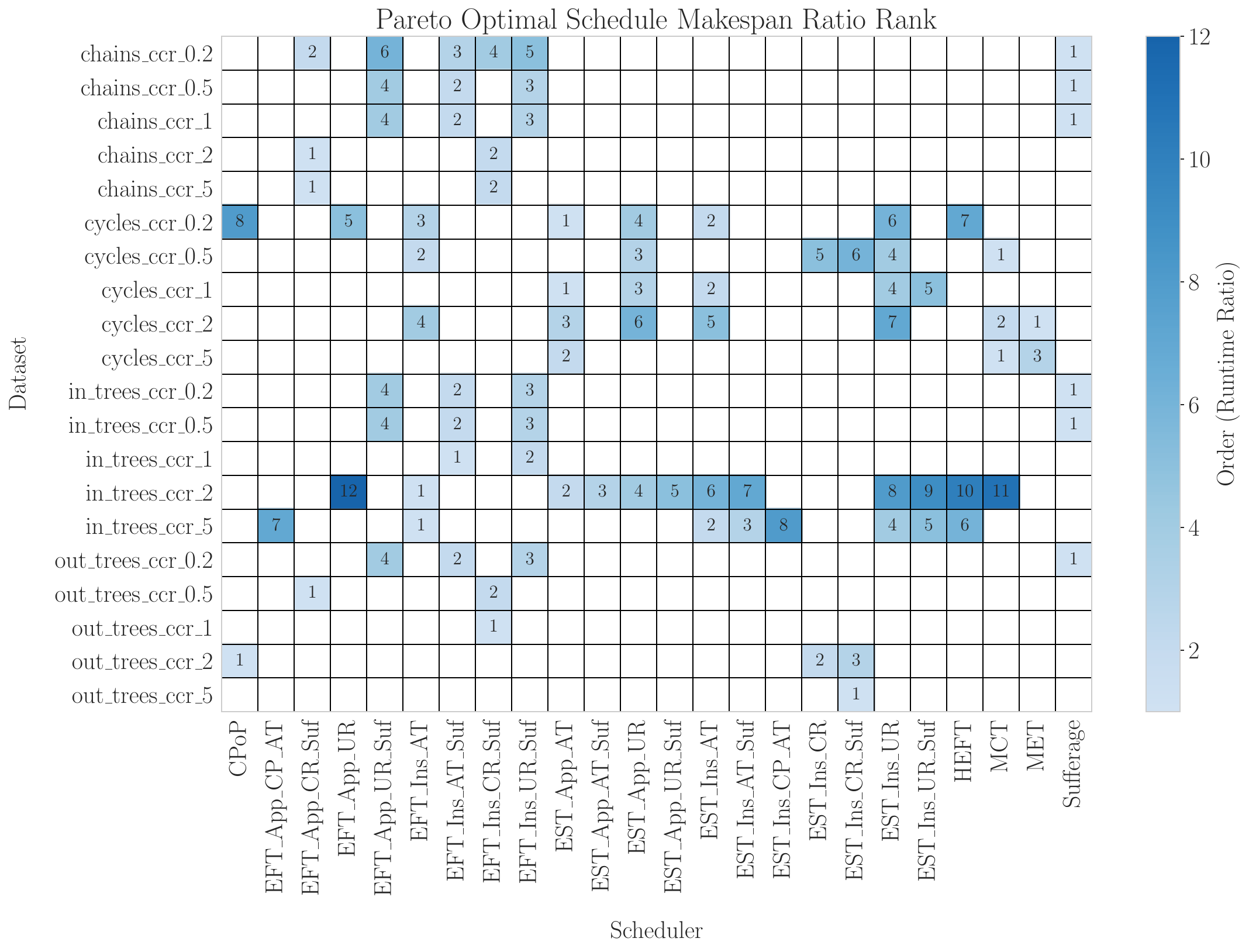}
        \caption{The chart number represents the pareto-optimal schedule's position (from left to right) for each dataset.}\label{fig:pareto:chart}
    \end{subfigure}
    \caption{Pareto-optimal scheduling algorithms for each dataset.}
    \label{fig:pareto}
\end{figure*}

A pareto-optimal scheduler is one such that no other scheduler has \textit{both} a lower average makespan ratio \textit{and} a lower average runtime ratio on a given dataset.
Figure~\ref{fig:pareto:scatter} depicts the pareto-optimal schedulers for each dataset.
Each subplot has $24$ markers (one for each pareto-optimal scheduler) but only the schedulers that are pareto optimal for each dataset are colored blue.
For example, for the dataset in\_trees\_ccr\_0.2 (in\_trees with CCR=1/5), two schedulers that are pareto-optimal for another dataset are strictly dominated by another scheduler for this dataset (and so are colored red).
Figure~\ref{fig:pareto:chart} serves as a kind of legend for the scatter plot.
Each cell represents the pareto-optimal scheduler's rank compared to other pareto-optimal schedulers for the same dataset with respect to its runtime ratio.
For each dataset, the scheduler labeled $1$ is that with the least runtime ratio (furthest blue marker to the left in the corresponding scatter plot in Figure~\ref{fig:pareto:scatter}).
Because it is pareto-optimal, this also indicates it has the \textit{highest} makespan ratio of pareto-optimal schedulers for the dataset (and so is also the \textit{highest} blue marker in the corresponding scatter plot in Figure~\ref{fig:pareto:scatter}).
Blank cells indicate that the scheduling algorithm is not pareto-optimal for the dataset.
Thus, schedulers that have consistently low order numbers (like Sufferage) are pareto-optimal mostly (or entirely) due to their low runtimes and \textit{not} due to their low makespan ratios.
Results indicate that these are scheduling algorithms that are fast, but not very performant.
On the other hand, scheduling algorithms that have consistently high order numbers (like HEFT) are pareto-optimal mostly (or entirely) due to their low makespan ratios despite higher runtimes.
These algorithms are performant, but slower in generating schedules.

\subsection{Effects of Algorithmic Components}
\begin{figure}[!h]
    \begin{subfigure}{0.5\linewidth}
        \centering
        \includegraphics[width=\linewidth]{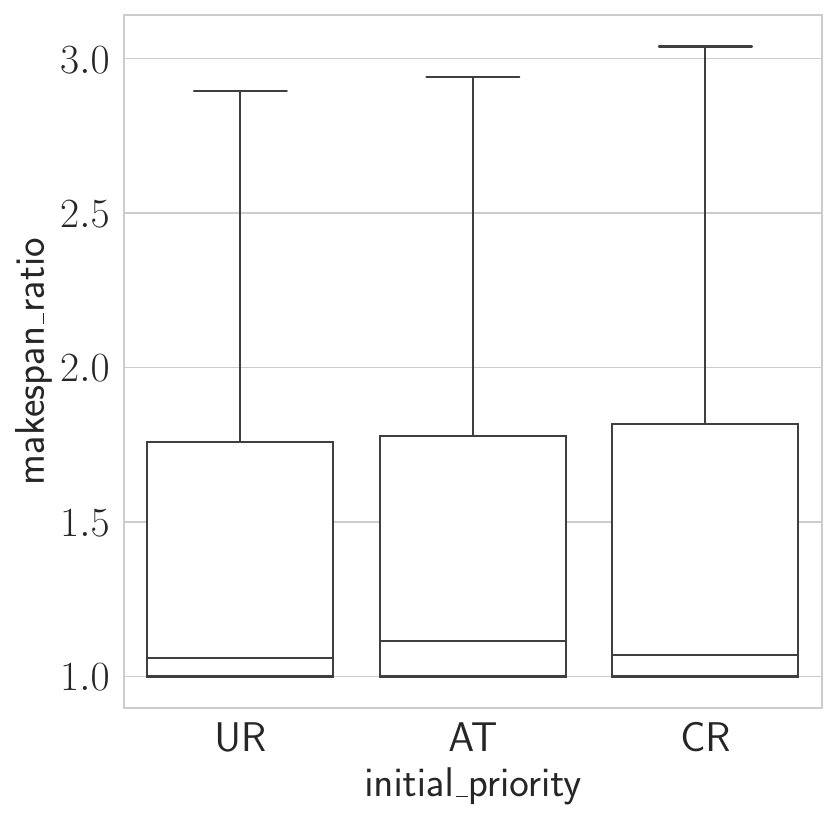}
    \end{subfigure}%
    \begin{subfigure}{0.5\linewidth}
        \centering
        \includegraphics[width=\linewidth]{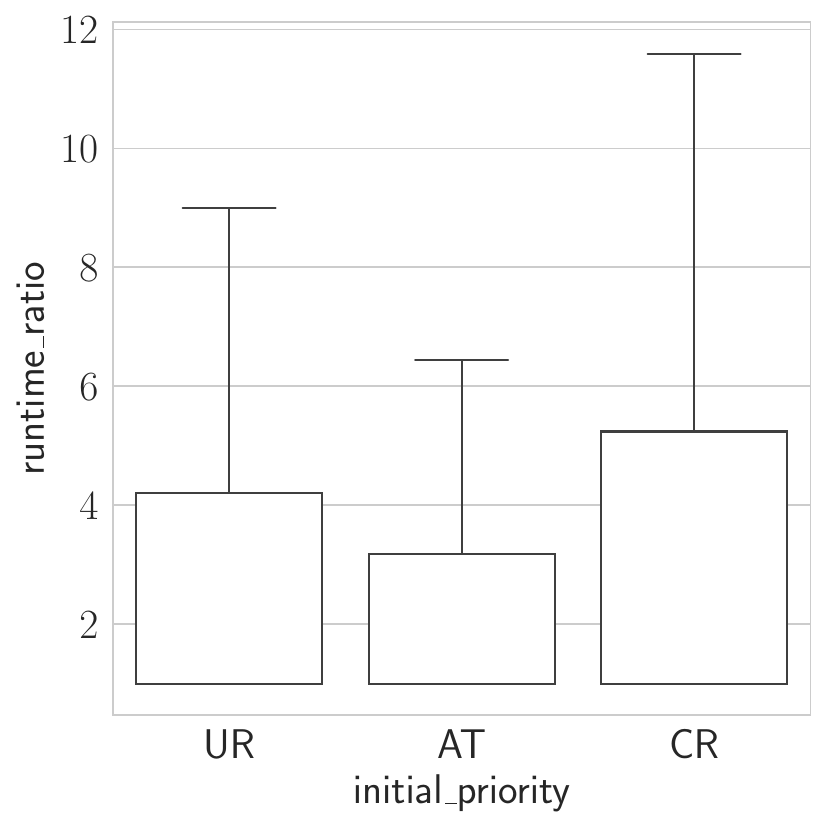}
    \end{subfigure}
    \caption{Effect of the \textit{initial priority function} on the makespan and runtime ratios over all datasets. UR stands for UpwardRanking, AT for ArbitraryTopological, and CR for CPoPRanking.}
    \label{fig:initial_priority}
\end{figure}

\begin{figure}[!h]
    \begin{subfigure}{0.5\linewidth}
        \centering
        \includegraphics[width=\linewidth]{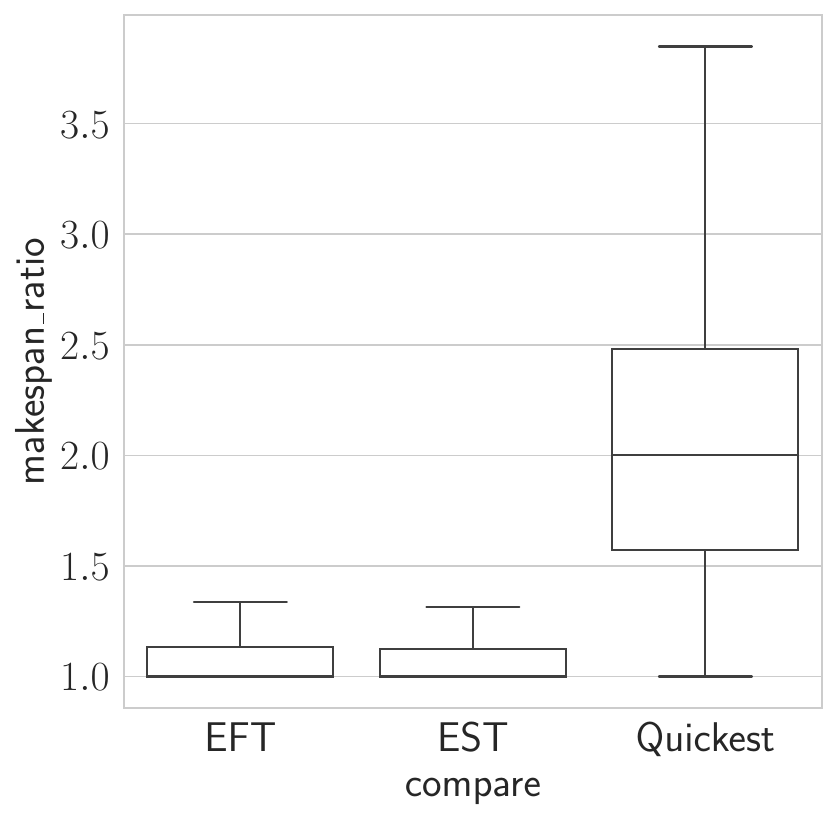}
    \end{subfigure}%
    \begin{subfigure}{0.5\linewidth}
        \centering
        \includegraphics[width=\linewidth]{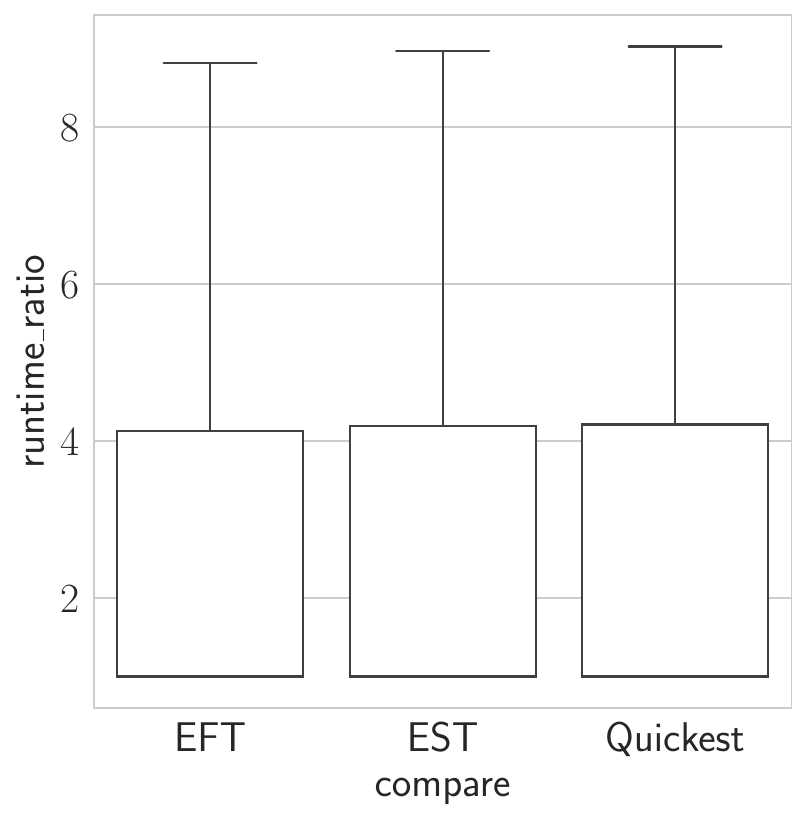}
    \end{subfigure}
    \caption{Effect of the \textit{comparison function} on the makespan and runtime ratios over all datasets.}
    \label{fig:compare}
\end{figure}

\begin{figure}[!h]
    \begin{subfigure}{0.5\linewidth}
        \centering
        \includegraphics[width=\linewidth]{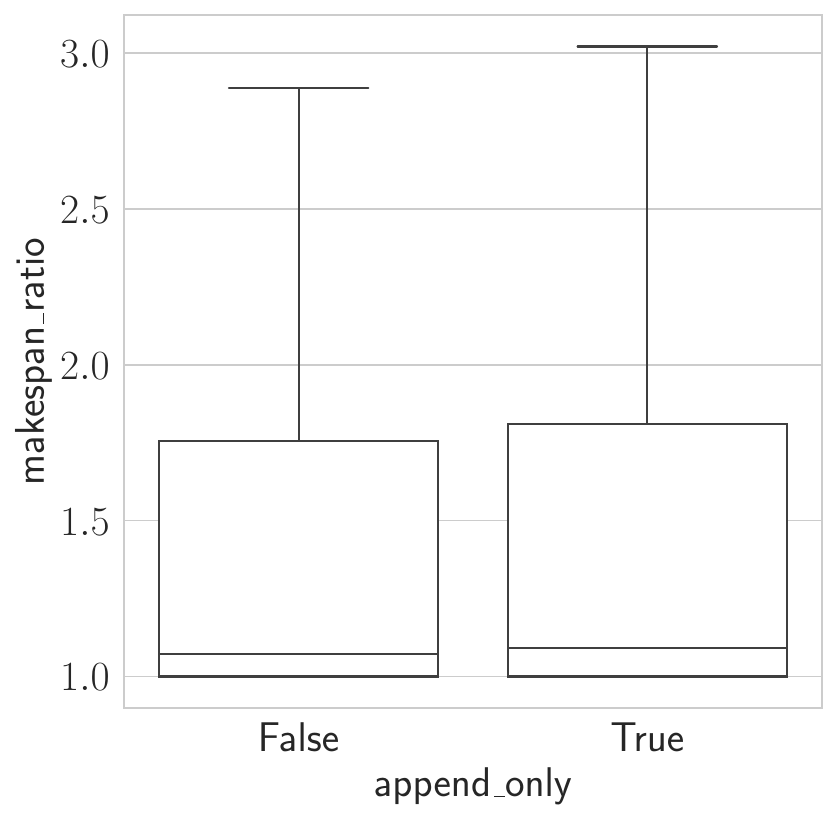}
    \end{subfigure}%
    \begin{subfigure}{0.5\linewidth}
        \centering
        \includegraphics[width=\linewidth]{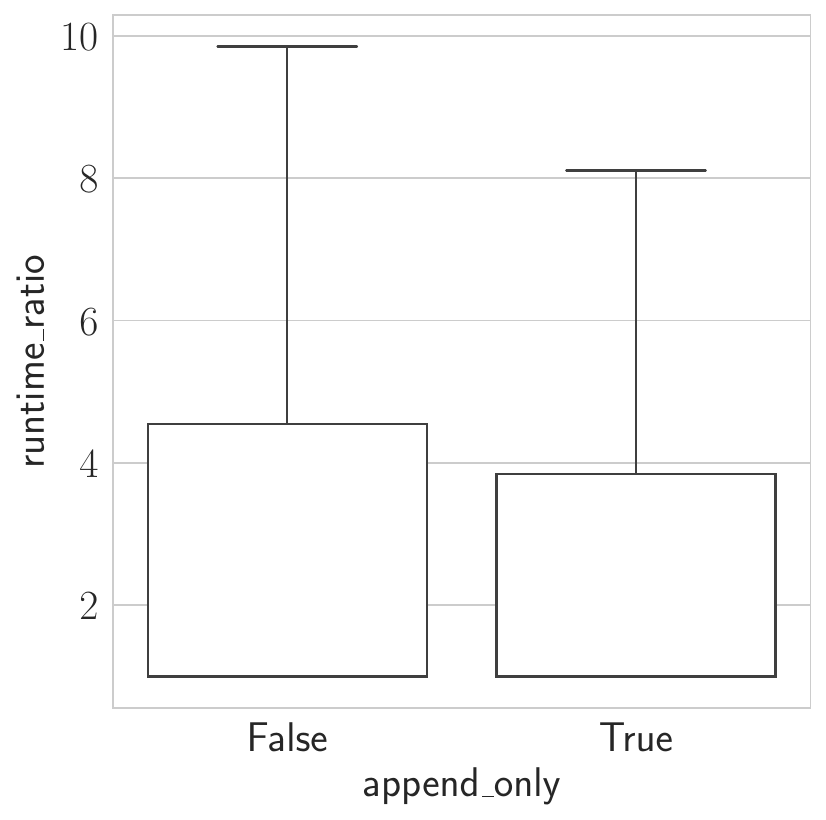}
    \end{subfigure}
    \caption{Effect of the \textit{insertion vs. append-only scheme} on the makespan and runtime ratios over all datasets.}
    \label{fig:append_only}
\end{figure}

\begin{figure}[!h]
    \begin{subfigure}{0.5\linewidth}
        \centering
        \includegraphics[width=\linewidth]{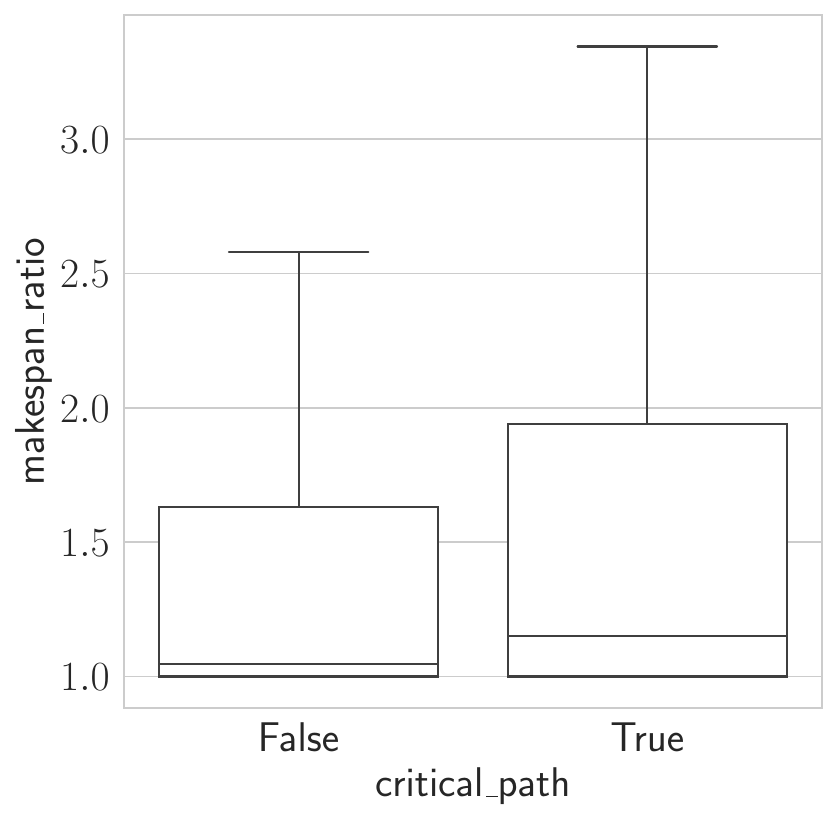}
    \end{subfigure}%
    \begin{subfigure}{0.5\linewidth}
        \centering
        \includegraphics[width=\linewidth]{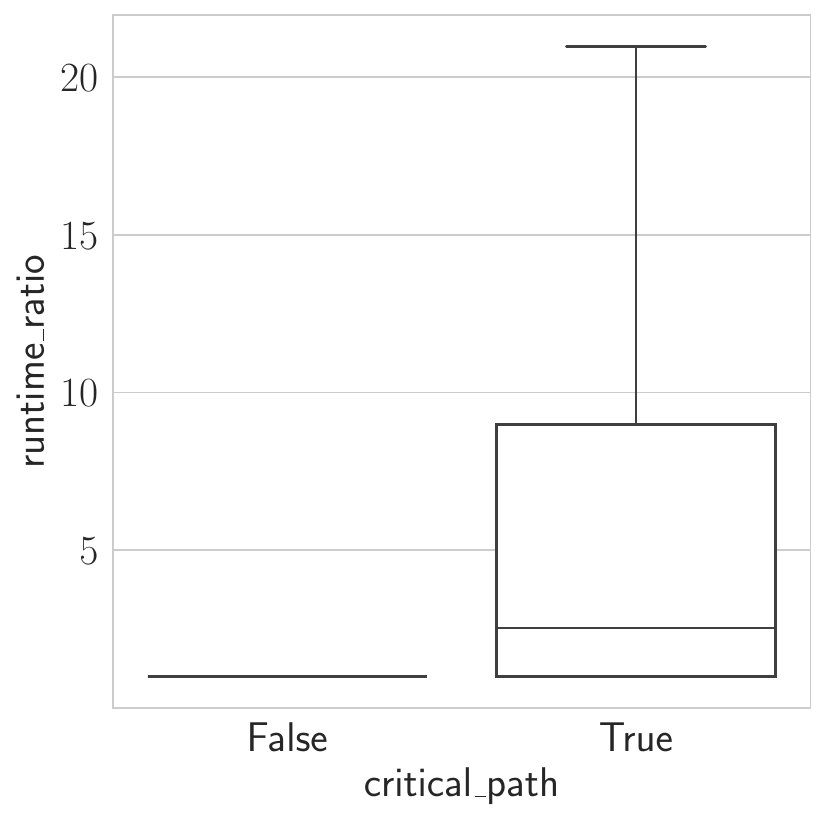}
    \end{subfigure}
    \caption{Effect of the \textit{critical path reservation} on the makespan and runtime ratios over all datasets.}
    \label{fig:critical_path}
\end{figure}

\begin{figure}[!h]
    \begin{subfigure}{0.5\linewidth}
        \centering
        \includegraphics[width=\linewidth]{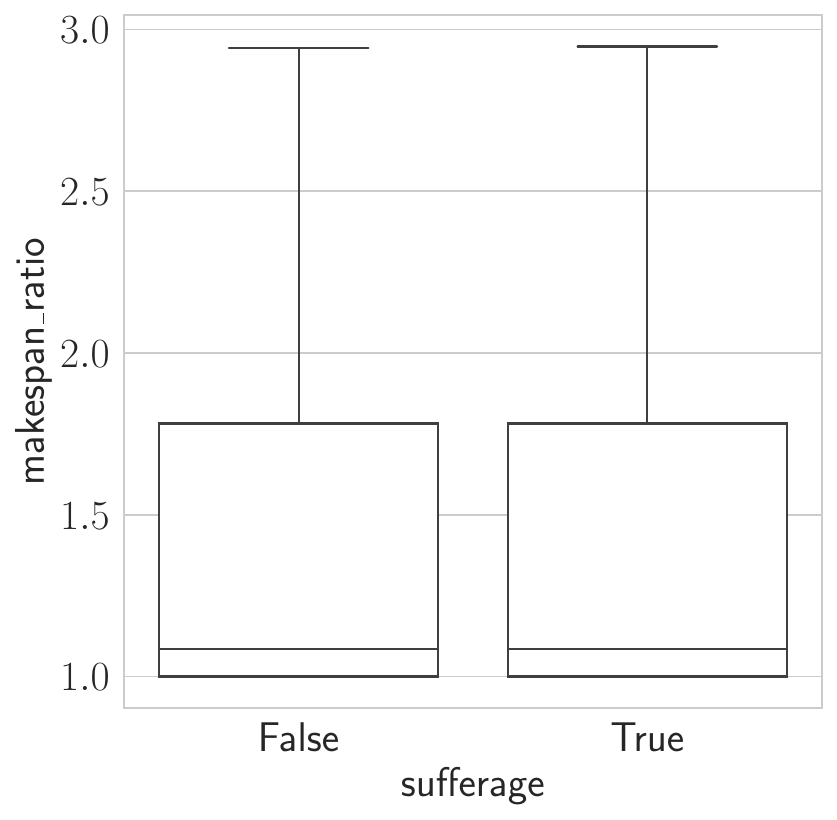}
    \end{subfigure}%
    \begin{subfigure}{0.5\linewidth}
        \centering
        \includegraphics[width=\linewidth]{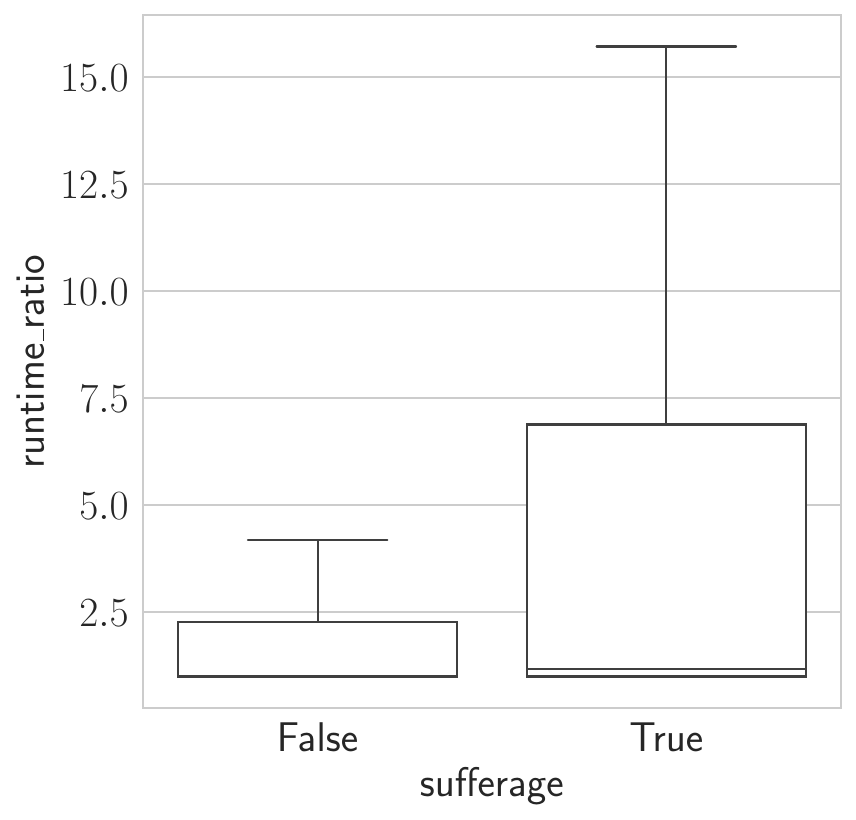}
    \end{subfigure}
    \caption{Effect of the \textit{sufferage selection scheme} on the makespan and runtime ratios over all datasets.}
    \label{fig:sufferage}
\end{figure}

In this section, we study the effect that different algorithmic components have on performance and runtime.
Figure~\ref{fig:initial_priority} suggests that, across all datasets, the priority function has a small effect on makespan ratio with UpwardRanking just slightly out-performing ArbitraryTopological and CPoPRanking.
Similar results for the append-only and sufferage components are shown in Figures~\ref{fig:append_only} and~\ref{fig:sufferage}.
Figure~\ref{fig:compare} indicates that while the Quickest comparison function is clearly the least performant, EFT and EST have roughly similar performance across all datasets.
Figure~\ref{fig:critical_path} shows that critical path reservation tends to increase makespan ratios and, more dramatically, increase runtime ratios, suggesting that critical path reservation, in general, is a poor strategy.
Recall, however, that Table~\ref{tab:pareto} from the previous section though, does indicate that there are datasets for which critical path reserving schedulers are pareto-optimal.
This suggests that the effects of these components might be dataset-specific.

\begin{figure}[!h]
    \begin{subfigure}{0.5\linewidth}
        \centering
        \includegraphics[width=\linewidth]{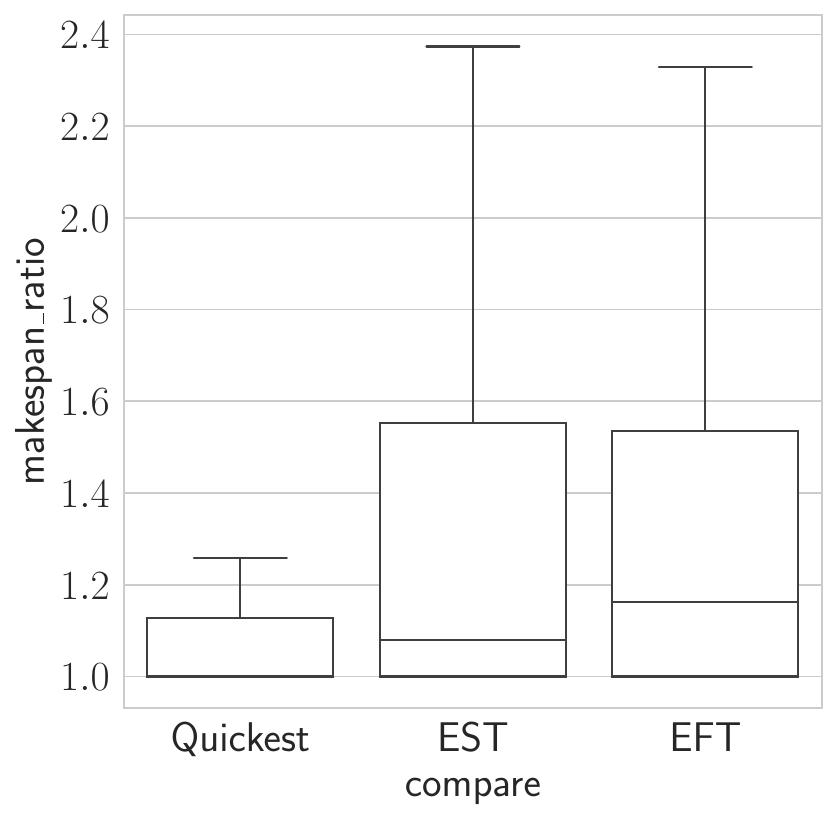}
    \end{subfigure}%
    \begin{subfigure}{0.5\linewidth}
        \centering
        \includegraphics[width=\linewidth]{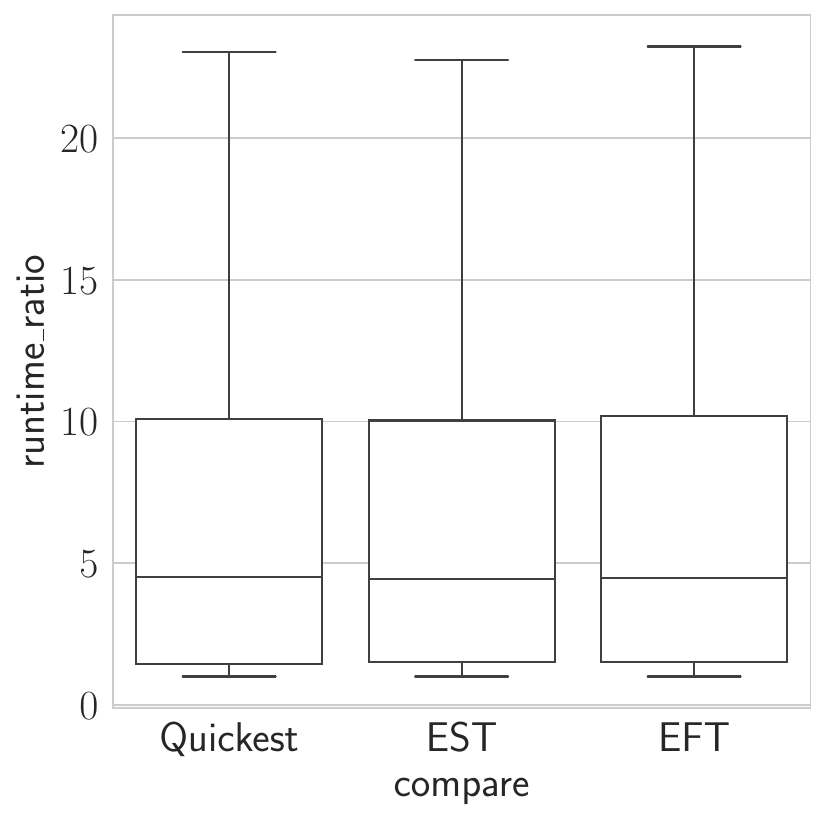}
    \end{subfigure}
    \caption{Effect of the \textit{comparison function} on the makespan and runtime ratios for the cycles\_ccr\_5 dataset (cycles task graphs with CCR=5).}
    \label{fig:compare:cycles_ccr_5}
\end{figure}

\begin{figure}
    \begin{subfigure}{\linewidth}
        \centering
        \includegraphics[width=\linewidth]{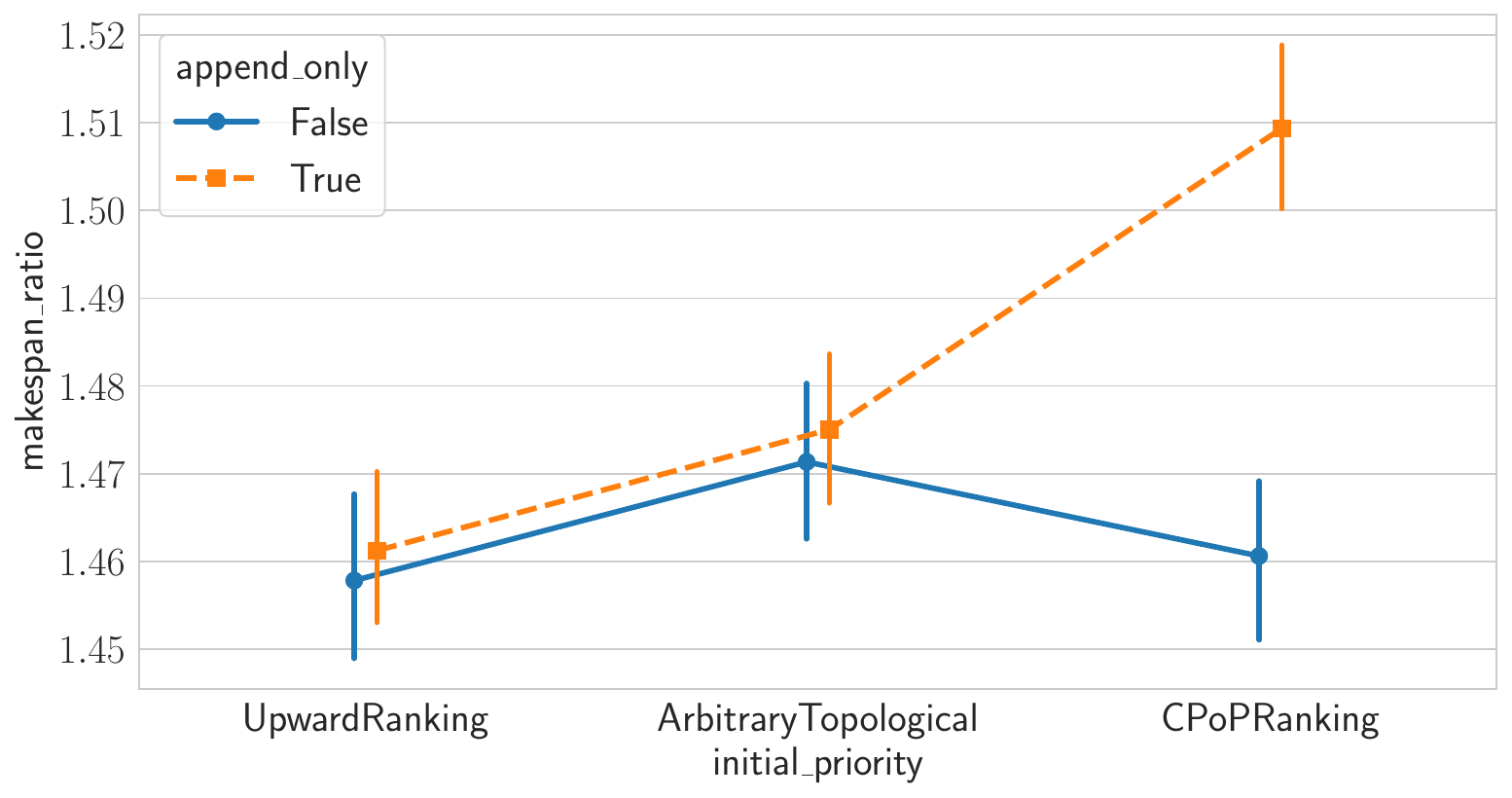}
        \caption{Interaction between append\_only and priority parameters.}\label{fig:int:priority:append}
    \end{subfigure}%

    \vspace{0.2cm}
    
    \begin{subfigure}{\linewidth}
        \centering
        \includegraphics[width=\linewidth]{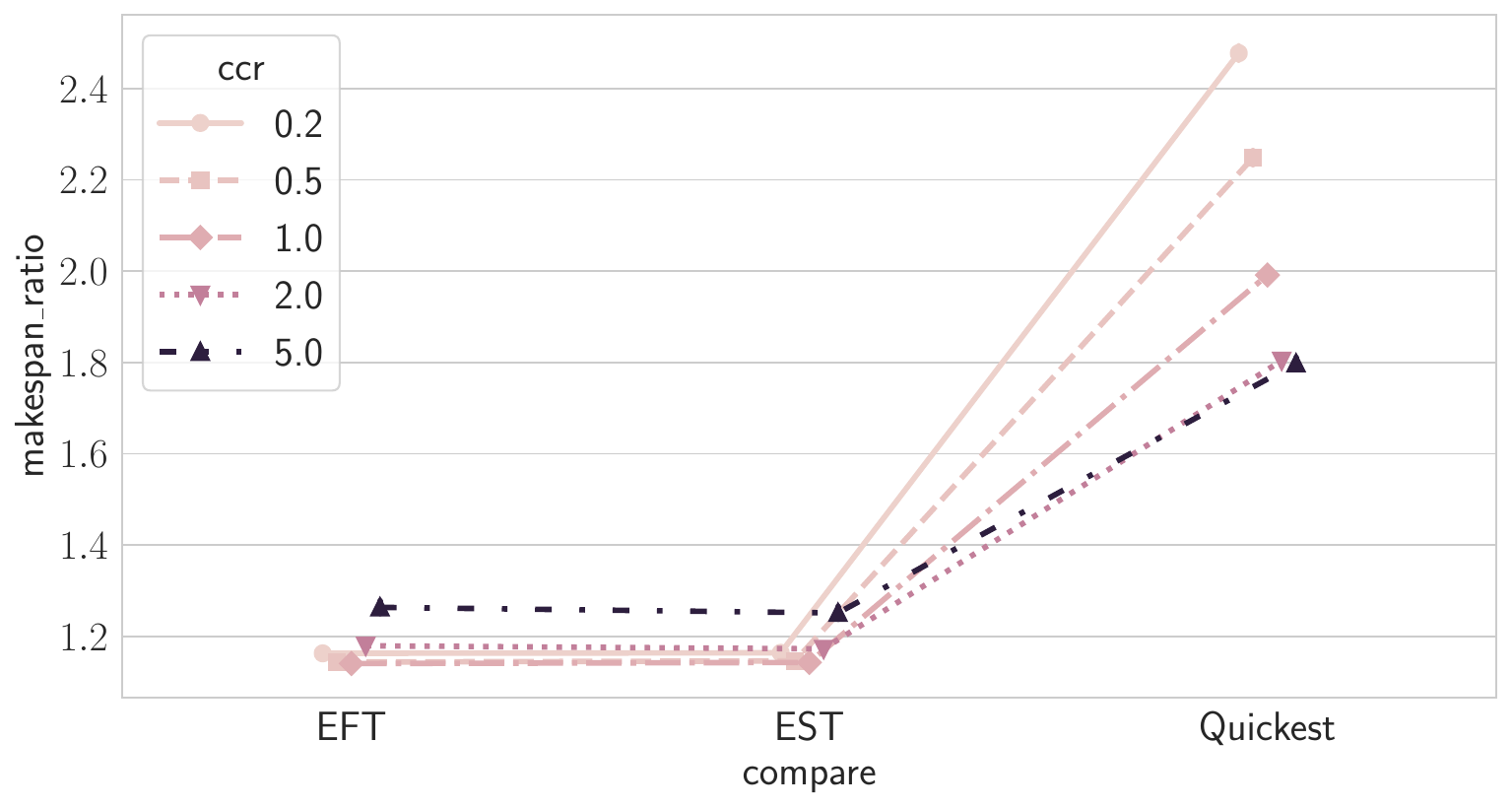}
        \caption{Interaction between the compare paremeter and the task graph's CCRs.}\label{fig:int:compare:ccr}
    \end{subfigure}
    
    \vspace{0.2cm}
    
    \begin{subfigure}{\linewidth}
        \centering
        \includegraphics[width=\linewidth]{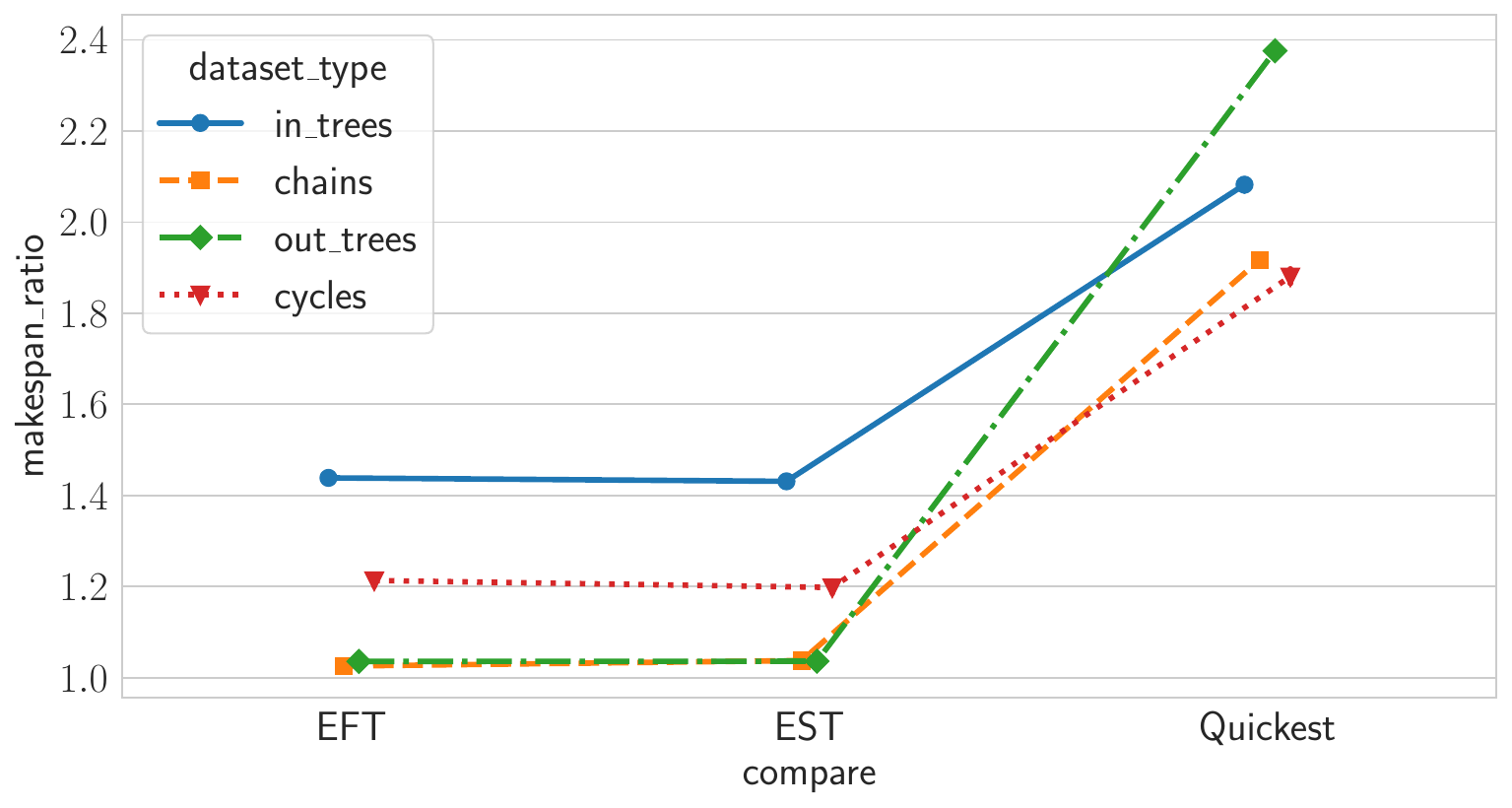}
        \caption{Interaction between compare function parameter and the dataset type.}\label{fig:int:compare:dataset_type}
    \end{subfigure}%

    \vspace{0.2cm}
    
    \begin{subfigure}{\linewidth}
        \centering
        \includegraphics[width=\linewidth]{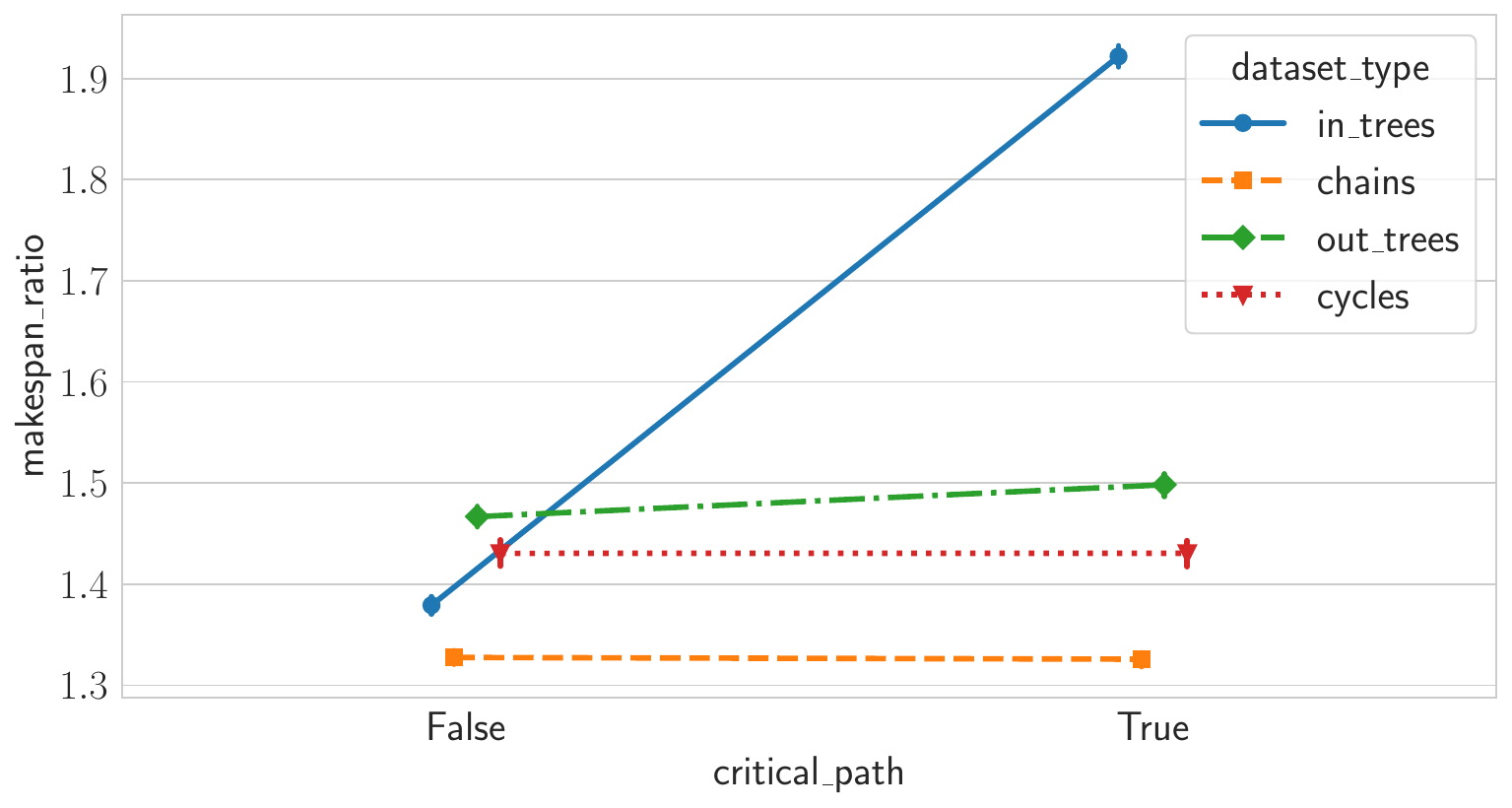}
        \caption{Interaction between the critical\_path parameter and the dataset type.}\label{fig:int:critical_path:dataset_type}
    \end{subfigure}
    \caption{Interactions Between Algorithmic Components}
    \label{fig:interactions}
\end{figure}

We do, in fact, observe in some cases quite different behavior for individual datasets.
For example, Figure~\ref{fig:compare:cycles_ccr_5} depicts the effect of the comparison function on the makespan and runtime ratios for the cycles\_ccr\_5 dataset (cycles task graphs with CCR=5).
The Quickest comparison function, which performs generally terribly compared to EFT and EST for other datasets, outperforms EFT and EST by a large margin!
The dataset isn't the only thing that interacts with the effects of algorithmic components, though.
Figure~\ref{fig:interactions} depicts a few of the more interesting \textit{interactions} between algorithmic components (averaged over all datasets).
Figure~\ref{fig:int:priority:append} depicts an interaction between the append\_only and initial\_priority parameters, suggesting that the combination of append\_only=True and intial\_priority=CPoPRanking has a more detrimental effect on the makespan ratio that either of the parameters do by themselves.
In other words, the append-only strategy is particularly bad when using the CPoPRanking priority function.
Figure~\ref{fig:int:compare:ccr} shows that the Quickest comparison function is generally bad but less so on communication-heavy applications (those with high CCR).
Figure~\ref{fig:int:compare:dataset_type} shows that it is also a particularly bad comparison function for the out\_trees datasets.
Figure~\ref{fig:int:critical_path:dataset_type} suggests that the small difference in the makespan ratio for schedulers that do critical path reservation may be due almost entirely to how critical path reservation increases makespan ratios for in\_trees datasets.
Plots for the individual effects and interactions between algorithmic components can be found in the appendix.

\section{Conclusion}\label{sec:conclusion}
In this paper, we proposed a generalized parametric list scheduling approach for studying the individual and combined effects of different algorithmic components.
We evaluated 72 algorithms produced by combining five different algorithmic components on 20 datasets and presented results on the individual and combined effects of these components on average performance and runtime across all datasets.
We also discuss how these results differ for individual datasets, suggesting that the way algorithmic components interact with each other is problem-dependent (i.e., depends on the task graph structure, whether or not the application is communication or computation heavy, etc.).
This work opens many avenues for future research.
First, this work can be extended by considering new algorithmic components (e.g., k-depth lookahead), new implementations for the five current algorithmic components, and other datasets.
In particular, it would be interesting to see more results for application-specific datasets like cycles.

In this paper, we compared algorithms using the traditional benchmarking approach, whereby we ran each scheduler on different datasets.
It has been shown that this approach, while certainly useful, can be misleading in some instances~\cite{pisa}.
An adversarial approach to comparing algorithms was recently proposed to address this.
It may be interesting to evaluate the scheduling algorithms and algorithmic components using this approach.

\section*{Acknowledgements}
This work was supported in part by Army Research Laboratory under Cooperative Agreement W911NF-17-2-0196.
The authors acknowledge the Center for Advanced Research Computing (CARC) at the University of Southern California for providing computing resources that have contributed to the research results reported within this publication. URL: \url{https://carc.usc.edu}.

\FloatBarrier
\bibliographystyle{IEEEtran}
\bibliography{main}

\begin{thebibliography}{10}
\providecommand{\url}[1]{#1}
\csname url@samestyle\endcsname
\providecommand{\newblock}{\relax}
\providecommand{\bibinfo}[2]{#2}
\providecommand{\BIBentrySTDinterwordspacing}{\spaceskip=0pt\relax}
\providecommand{\BIBentryALTinterwordstretchfactor}{4}
\providecommand{\BIBentryALTinterwordspacing}{\spaceskip=\fontdimen2\font plus
\BIBentryALTinterwordstretchfactor\fontdimen3\font minus \fontdimen4\font\relax}
\providecommand{\BIBforeignlanguage}[2]{{%
\expandafter\ifx\csname l@#1\endcsname\relax
\typeout{** WARNING: IEEEtran.bst: No hyphenation pattern has been}%
\typeout{** loaded for the language `#1'. Using the pattern for}%
\typeout{** the default language instead.}%
\else
\language=\csname l@#1\endcsname
\fi
#2}}
\providecommand{\BIBdecl}{\relax}
\BIBdecl

\bibitem{theory:npcomplete}
M.~R. Garey and D.~S. Johnson, \emph{Computers and Intractability: {A} Guide to the Theory of NP-Completeness}.\hskip 1em plus 0.5em minus 0.4em\relax W. H. Freeman, 1979.

\bibitem{theory:inapproximable}
\BIBentryALTinterwordspacing
A.~Bazzi and A.~Norouzi{-}Fard, ``Towards tight lower bounds for scheduling problems,'' in \emph{Algorithms - {ESA} 2015 - 23rd Annual European Symposium, Patras, Greece, September 14-16, 2015, Proceedings}, ser. Lecture Notes in Computer Science, N.~Bansal and I.~Finocchi, Eds., vol. 9294.\hskip 1em plus 0.5em minus 0.4em\relax Springer, 2015, pp. 118--129. [Online]. Available: \url{https://doi.org/10.1007/978-3-662-48350-3\_11}
\BIBentrySTDinterwordspacing

\bibitem{framework_repo}
\BIBentryALTinterwordspacing
A.~Authors, ``Scheduling algorithms gathered,'' Github, 2023. [Online]. Available: \url{https://anonymous.4open.science/r/saga-1F6D/README.md}
\BIBentrySTDinterwordspacing

\bibitem{graham}
\BIBentryALTinterwordspacing
R.~L. Graham, ``Bounds on multiprocessing timing anomalies,'' \emph{SIAM Journal on Applied Mathematics}, vol.~17, no.~2, pp. 416--429, 1969. [Online]. Available: \url{https://doi.org/10.1137/0117039}
\BIBentrySTDinterwordspacing

\bibitem{scheduler:heft}
\BIBentryALTinterwordspacing
H.~Topcuoglu, S.~Hariri, and M.~Wu, ``Task scheduling algorithms for heterogeneous processors,'' in \emph{8th Heterogeneous Computing Workshop, {HCW} 1999, San Juan, Puerto Rico, April12, 1999}.\hskip 1em plus 0.5em minus 0.4em\relax {IEEE} Computer Society, 1999, pp. 3--14. [Online]. Available: \url{https://doi.org/10.1109/HCW.1999.765092}
\BIBentrySTDinterwordspacing

\bibitem{compare:list_vs_cluster}
\BIBentryALTinterwordspacing
H.~Wang and O.~Sinnen, ``List-scheduling versus cluster-scheduling,'' \emph{{IEEE} Trans. Parallel Distributed Syst.}, vol.~29, no.~8, pp. 1736--1749, 2018. [Online]. Available: \url{https://doi.org/10.1109/TPDS.2018.2808959}
\BIBentrySTDinterwordspacing

\bibitem{compare:benchmarking_hetero}
\BIBentryALTinterwordspacing
A.~K. Maurya and A.~K. Tripathi, ``On benchmarking task scheduling algorithms for heterogeneous computing systems,'' \emph{J. Supercomput.}, vol.~74, no.~7, pp. 3039--3070, 2018. [Online]. Available: \url{https://doi.org/10.1007/s11227-018-2355-0}
\BIBentrySTDinterwordspacing

\bibitem{compare:benchmarking_schedulers}
Y.-K. Kwok and I.~Ahmad, ``Benchmarking the task graph scheduling algorithms,'' in \emph{Proceedings of the First Merged International Parallel Processing Symposium and Symposium on Parallel and Distributed Processing}, 1998, pp. 531--537.

\bibitem{scheduler:eleven}
\BIBentryALTinterwordspacing
T.~D. Braun, H.~J. Siegel, N.~Beck, L.~B{\"{o}}l{\"{o}}ni, M.~Maheswaran, A.~I. Reuther, J.~P. Robertson, M.~D. Theys, B.~Yao, D.~A. Hensgen, and R.~F. Freund, ``A comparison of eleven static heuristics for mapping a class of independent tasks onto heterogeneous distributed computing systems,'' \emph{J. Parallel Distributed Comput.}, vol.~61, no.~6, pp. 810--837, 2001. [Online]. Available: \url{https://doi.org/10.1006/jpdc.2000.1714}
\BIBentrySTDinterwordspacing

\bibitem{compare:metaheuristics}
T.~Braun, H.~Siegal, N.~Beck, L.~Boloni, M.~Maheswaran, A.~Reuther, J.~Robertson, M.~Theys, B.~Yao, D.~Hensgen, and R.~Freund, ``A comparison study of static mapping heuristics for a class of meta-tasks on heterogeneous computing systems,'' in \emph{Proceedings. Eighth Heterogeneous Computing Workshop (HCW'99)}, 1999, pp. 15--29.

\bibitem{scheduler:sufferage}
\BIBentryALTinterwordspacing
T.~N'Takp{\'{e}} and F.~Suter, ``Critical path and area based scheduling of parallel task graphs on heterogeneous platforms,'' in \emph{12th International Conference on Parallel and Distributed Systems, {ICPADS} 2006, Minneapolis, Minnesota, USA, July 12-15, 2006}.\hskip 1em plus 0.5em minus 0.4em\relax {IEEE} Computer Society, 2006, pp. 3--10. [Online]. Available: \url{https://doi.org/10.1109/ICPADS.2006.32}
\BIBentrySTDinterwordspacing

\bibitem{data:random_graphs}
D.~Cordeiro, G.~Mouni\^{e}, S.~Perarnau, D.~Trystram, J.-M. Vincent, and F.~Wagner, ``Random graph generation for scheduling simulations.''\hskip 1em plus 0.5em minus 0.4em\relax ICST, 5 2010.

\bibitem{data:cycles}
\BIBentryALTinterwordspacing
R.~F. da~Silva, R.~Mayani, Y.~Shi, A.~R. Kemanian, M.~Rynge, and E.~Deelman, ``Empowering agroecosystem modeling with {HTC} scientific workflows: The cycles model use case,'' in \emph{2019 {IEEE} International Conference on Big Data {(IEEE} BigData), Los Angeles, CA, USA, December 9-12, 2019}, C.~K. Baru, J.~Huan, L.~Khan, X.~Hu, R.~Ak, Y.~Tian, R.~S. Barga, C.~Zaniolo, K.~Lee, and Y.~F. Ye, Eds.\hskip 1em plus 0.5em minus 0.4em\relax {IEEE}, 2019, pp. 4545--4552. [Online]. Available: \url{https://doi.org/10.1109/BigData47090.2019.9006107}
\BIBentrySTDinterwordspacing

\bibitem{pisa}
J.~Coleman and B.~Krishnamachari, ``Comparing task graph scheduling algorithms: An adversarial approach,'' under review.

\end{thebibliography}

\end{document}